\documentclass{IEEEtaes}
\IEEEoverridecommandlockouts
\usepackage{amsmath,amsfonts}
\usepackage{algpseudocode}
\usepackage{array}
\usepackage{textcomp}
\usepackage{stfloats}
\usepackage{url}
\usepackage{verbatim}
\usepackage{graphicx}
\usepackage{cite}
\usepackage{amssymb}
\usepackage{bm}
\usepackage{amsthm}
\usepackage[thinc]{esdiff}
\usepackage{xcolor}
\usepackage{float}
\usepackage{caption}
\usepackage{subcaption}
\usepackage[utf8]{inputenc}
\usepackage[linesnumbered, ruled]{algorithm2e}
\SetKwRepeat{Do}{do}{while}%
\newtheorem{prop}{Proposition}

\newtheorem{theorem}{Theorem}

\begin{document}

\title{Cooperative Sense and Avoid for UAVs using Secondary Radar}

\author{Mostafa Mohammadkarimi}
\member{Member, IEEE}
\affil{Delft University of Technology, Delft, The Netherlands}

\author{Raj Thilak Rajan}
\member{Senior Member, IEEE}
\affil{Delft University of Technology, Delft, The Netherlands}


\receiveddate{This work is partially funded by the European Leadership Joint Undertaking (ECSEL JU), under grant agreement No 876019, and the ADACORSA project - ``Airborne Data Collection on Resilient System Architectures.” (https://adacorsa.eu/). }

\authoraddress{
The authors are with the Faculty of Electrical Engineering,
Mathematics, and Computer Science, Delft University of Technology, 2628 CD
Delft, The Netherlands (e-mail: m.mohammadkarimi@tudelft.nl, R.T.Rajan@tudelft.nl).}

\authoraddress{This paper has been submitted to IEEE Transactions on Aerospace and Electronic Systems}

\maketitle

\begin{abstract}
A cooperative Sense and Avoid (SAA) algorithm for safe navigation of small-sized UAVs within an airspace is proposed in this paper.
The proposed method relies upon cooperation between the UAV and the surrounding transponder-equipped aviation obstacles.
To do so, the aviation obstacles
share their altitude and identification code with the UAV by using a Mode S operation of the Secondary Surveillance Radar (SSR) after interrogation.
The proposed SAA algorithm benefits from the estimate of the aviation obstacle's elevation angle for ranging. This results in more accurate ranging compared to the round-trip time-based ranging, which is currently used in existing SAA systems.
We also propose a low-complexity and accurate radial velocity estimator for the Mode S operation of the SSR which is employed in the proposed SAA system. Furthermore, by considering the Pulse-Position Modulation (PPM) of the transponder reply as a waveform of pulse radar with
random pulse repetition intervals, the maximum unambiguous radial velocity is obtained.
The proposed SAA is equipped with an intruder identification method that determines the risk level of the surrounding transponder-equipped aviation obstacles.
Given the estimated parameters, the intruder identification method classifies the aviation obstacles into high- medium-, and low-risk intruders. The output of the classifier enables the UAV to plan its path or maneuver for safe navigation accordingly. The
root mean square error (RMSE) of the proposed estimators are analytically derived, and the effectiveness of our SAA solution is confirmed through simulation experiments.

\begin{IEEEkeywords}
Cooperative sense and avoid, UAV, Secondary Surveillance Radar, Mode S, transponder, Intruder identification, 2D-MUSIC, Doppler shift, Radial velocity.
\end{IEEEkeywords}
\end{abstract}

\section{Introduction}
Sense and avoid (SAA) systems enable manned aerial vehicles (MAVs) and unmanned aerial vehicles (UAVs)
to integrate safely into civilian airspace by avoiding collisions with other obstacles \cite{son2017brief,tomasello2011detect,fasano2016sense,mccalmont2002detect}.
A SAA system first monitors the environment surrounding the aerial vehicle (AV) by employing different types of onboard sensors.
Following which, it decides whether a collision is impendent and generates a new flight path in order to avoid collision. An advanced SAA system can take advantage of sensor fusion algorithms in combination with image recognition and artificial intelligence to increase the reliability and precision of the system \cite{lyu2018detect}. However, performance improvement is typically achieved at the expense of higher computational complexity and power consumption, which makes on-board processing infeasible for small- and moderate-sized UAVs \cite{yu2015sense}.

Existing SAA solutions for AV can be categorized into non-cooperative and cooperative systems. The former SAA systems do not require the co-operation of the obstacles and the AV in the SAA procedure. The sensors used in these SAA systems can be active and/or passive.
Active sensors, such as primary radar, ultrasound, and LiDAR, emit a signal that is then reflected by an obstacle and detected again by the sensor \cite{nijsure2016cognitive,ben2017evaluation,aldao2022lidar}. Passive sensors detect a signal emitted by the object itself, and include visual and infrared cameras.
The later SAA systems
are based on the cooperation of the AV and the aviation obstacles (transponder-equipped AVs)
within an airspace.
The cooperation between the AV and aviation obstacles can be achieved through the Secondary Surveillance Radar (SSR) Mode A, C, and S transponders \cite{orlando1989mode}. Existing cooperative SAA systems mainly employ the SSR in its Mode S operation. These systems can be categorized as: 1) SAA with
an onboard interrogator, such as the Traffic Collision and Avoidance System (TCAS), internationally
known as the Airborne Collision Avoidance System (ACAS) \cite{smith2022understanding,berges2020feasibility,lin2015sense}, and 2) SAA without an onboard interrogator, for example,
the automatic dependent surveillance–broadcast (ADS-B) \cite{barnhart2021introduction,angelov2012sense,semke2017analysis,strohmeier2014realities}.
Note that the mode S transponders are compatible with Mode A and Mode C SSR.

The ACAS surveillance unit interrogates nearby aviation obstacles and then tracks them in the surrounding airspace through replies from their air traffic control transponders. These SAA systems typically obtain the range of each aviation obstacle by measuring the round-trip time (RTT) between the transmission of the Mode S interrogation and the receipt of the Mode S downlink reply \cite{kuchar2007traffic}. The altitude and vertical velocity of each aviation obstacle
are determined by tracking the response information of the downlink Mode S \cite{sahawneh2015detect}.
In the SAA systems using ADS-B, the AV and the aviation obstacles are equipped with Mode S transponder and either periodically broadcast and receive identification, position and other state information without Mode S interrogation.
The exchange of position information by the AV and aviation obstacles necessitates the need of Global Positioning System  in the ADS-B based SAA systems.

\begin{table}\label{nonumber}
\centering
\caption{{List of acronyms and symbols}}
\begin{tabular}{ |l|l| }
  \hline
  ACAS & Airborne Collision Avoidance System \\
  ADC & analog-to-digital converter \\
  AOA-E & angle-of-arrival estimator \\
  AV & aerial vehicle \\
  AWGN & additive white Gaussian noise \\
  CRC & cyclic redundancy check \\
  MAV & manned aerial vehicle \\
  MUSIC & MUltiple SIgnal Classification \\
  NC-SD & noncoherent symbol detector \\
  PPM & pulse-position modulation \\
  PSD & power spectral density \\
  PRI & pulse repetition interval \\
  RF & radio frequency \\
  RMSE & root mean square error \\
  RTT & round-trip time \\
  RV-E & radial velocity estimator \\
  SAA & sense and avoid  \\
  SSR & secondary surveillance radar \\
  S2V & Sequence to Vector \\
  TCAS & Traffic Collision Avoidance System \\
  TO-E & timing offset estimator \\
  UAV & unmanned aerial vehicle \\
  V2M & Vector to Matrix \\
  2D & two-dimensional \\
  $\phi$ & azimuth angle of an aviation obstacle \\
  $\theta$ & elevation angle of an aviation obstacle \\
  $L$ & free space pathloss \\
  $R$ & range of the aviation obstacle from the UAV \\
  $T$ & half of the PPM symbol time  \\
  $T_{\rm s}$ & sampling time \\
  $B$ & bandwidth of the received filter \\
  ${\bf f}$ & carrier frequency offset vector (omnidirectional)    \\
  ${\bf f}^{\rm a}$ & carrier frequency offset vector (antenna array)    \\
  $h$ & altitude of the UAV measured by its altimeter \\
    $h_{\rm a}$ & altitude of the aviation obstacle  \\
  $N$ & number of observation samples   \\
  $N_{\rm a}$ & number of elements in the planar antenna array  \\
  $M_{\rm a}$ & number of  scans sweeping the azimuth angle \\
  $M_{\rm e}$ & number of scans sweeping the elevation angle \\
  $\tau$ &  time delay of the DF4 reply at the UAV \\
  $\tau_{\rm max}$ &  maximum time delay for the reception of DF4   \\
    $\tau_{\rm gr}$&  group delay of the receive low-pass filter    \\
  ${\bf r}$ &  DF4 preamble vector of length $16$  \\
  $\bar{\bf r}$ &  extended preamble vector  \\
  ${\bf b}$ &  DF4 payload vector of length $56$ \\
  $P_{\rm t}$ &  Peak transmit power of the DF4 reply  \\
  $f_{\rm D}$ &  Doppler frequency shift  \\
  $z(t)$ &  Mode S trapezoidal transmit pulse   \\
   $h(t-\tau_{\rm gr})$ &  low-pass filter at the receiver    \\
  $g(t)$ &  received filtered pulse   \\
$y(t)$ & received  baseband signal at the UAV antenna \\
  ${\bf g}$ & vector of length $M$ including the samples of $g(t)$ \\
  $v_{\rm r}$ & relative radial velocity   \\
  $M$ & number of samples per pulse   \\
  ${\bf q}$ & DF4 preamble and payload vector   \\
  $N_0$ & PSD of noise (watts per hertz)    \\ 
  $N_{\rm av}$ & number of detected aviation obstacles    \\
  $\bf y$ &  received vector by the omnidirectional antenna  \\
  ${\bf Y}^{\rm a}$ &  received matrix by the antenna array \\
   ${\bf w}$ &  AWGN vector of the omnidirectional antenna \\
  ${\bf W}^{\rm a}$ &  AWGN matrix of the antenna array  \\
    $\epsilon_{\theta}$ &  threshold for switching to the RTT-based ranging  \\
       $c$&  speed of light  \\
      $f_{\rm c}$&  carrier frequency of the DF4 reply \\ 
    $\lambda$&  wavelength of the DF4 reply \\  
    $\bf A$&  sample correlation matrix in the antenna array  \\     
  \hline
\end{tabular}\label{tabale:2}
\end{table}

Cooperative SAA offers a higher accuracy compared to the non-cooperative systems; however, it cannot detect non-aviation obstacles, such as buildings and power lines in urban landscapes. On the contrary, a fast and reliable detection and identification of small-sized UAVs by the means of non-cooperative SAA systems is
challenging. For example, typically primary radars cannot automatically classify
whether an object is a bird or a UAV because of the very low radar signatures of the small-sized UAVs (with a radar cross section of the order of $0.01 {\rm m}^2$) \cite{torvik2016classification}.
The key disadvantage of the non-cooperative SAA systems using active ultrasonic sensors or LiDAR is their short range sensing. Lastly, passive sensors, such as visible-light cameras, cannot capture images at night or in low light (at dusk or dawn, in fog, etc.), and detecting small-sized UAVs in infrared images is challenging \cite{rs7010001}.

Cooperative SAA is significantly effective for detecting and tracking small-sized UAVs.
 In this regard,
a promising solution is to employ
inertial navigation unit in combination with cooperative SAA equipped with an onboard Mode S interrogator and altimeter sensor for altitude measurement.

The radial velocity information of the aviation obstacles can improve the performance of a SAA system. Existing cooperative SAA with an onboard Mode S interrogator, such as
the ACAS, can estimate the radial velocity and the vertical velocity of the aviation obstacles by tracking the decoded altitude information, which requires the reception of at least two correctly decoded Mode S downlink replies from each aviation obstacle. Radial velocity estimation using altitude information is accurate for constant relative radial velocity, which is not the case in practice.
Furthermore, the standard
for Mode S responses allows small timing variances that, although small, affect system performance
when used to estimate RTT \cite{smith2022understanding}; hence,
the RTT-based ranging algorithm in the ACAS is not very accurate.

Taking into account the above-mentioned limitations and drawbacks of the ACAS,  we propose an efficient and low-complexity cooperative SAA system
for small-sized UAVs equipped with an onboard Mode S interrogator, inertial navigation unit, and an altimeter. The proposed SAA method can estimate the radial velocity of an aviation obstacle with a single Mode S downlink reply; hence, it can offer a more accurate radial velocity estimation as compared to the ACAS.
Moreover, it uses the decoded altitude information in the Mode S downlink reply and the estimated elevation angle of the aviation obstacles for ranging.
This can improve the accuracy of ranging compared to the RTT-based method used in the ACAS because ranging does not depend on the variances in timing of Mode S reply. Finally, based on sensing information, the proposed SAA method
classifies aviation obstacles as low-, medium-, and low-risk intruders. By employing the estimated parameters, the UAV can then decide on the best evasive maneuver or flight path correction to avoid collision.

{\subsection{Contributions:} Our key contributions are listed below.}
\begin{itemize}
\item We propose a cooperative SAA method with an onboard mode S interrogator by means of which an small-size UAV can estimate the radial velocity of the aviation obstacles in the airspace.
To do so, we propose a low-complexity and accurate estimator that estimates the Doppler frequency shift from the DF4 reply signal transmitted by the aviation obstacle. We analytically obtain the
root mean square error (RMSE)  of the proposed estimator and the maximum unambiguous radial velocity that can be estimated.

 \item The idea of multicarrier transmission for the separation of the Doppler frequency shift and the frequency drift between the oscillators of the aviation obstacle  and the UAV in the mode S operation SSR is proposed.

\item Range measurement by using the triangular relation between the elevation angle of the aviation obstacle  and the altitude difference is investigated for mode S operation SSR, and we analytically obtain the RMSE of the ranging method.

\item  A low-complexity and efficient intruder identification method for the mode S operation SSR is proposed.
\end{itemize}

{\subsection{Notation} Throughout the paper, the superscript ${\rm T}$} denotes the transpose operator, $*$ is the complex conjugation operator, the symbol $\neg$ is the logical not,
$[\cdot]$ and $\otimes$ denote the floor function and the Kronecker product, respectively. The convolution operator is shown by $\circledast$.
Vectors and matrices are denoted by bold lowercase
and uppercase letters, respectively, 
 and $\rm{diag}(\bf a)$
returns a diagonal matrix with entries of the vector $\bf a$ along the main diagonal. Vectors ${\bf{0}}_N$ and ${\bf{1}}_N$ denote the all-zero and all-ones vectors of length $N$, respectively,
and the symbol $\hat{{x}}$ is an estimate of $x$.
The continuous uniform distribution
between $a$ and $b$ is denoted by $\mathcal{U}_{\rm c}(a,b)$.
In Table I, a list of acronyms and the symbols used throughout the paper are listed.

\subsection{Outline}
The remaining of the paper is organized as follows. Section II introduces the system model. Section III describes the proposed SAA and
the signal model for the SAA system. In Section IV, the radial velocity estimation of the aviation obstacles is
discussed. The AOA estimation is studied in Section V. The proposed ranging algorithm and the intruder identification method are investigated in Sections VI and VII, respectively.
Simulation results are provided
in Section VIII, and conclusions are drawn in Section X.

\begin{figure}[!t]
\centering
\includegraphics[width=2.6in]{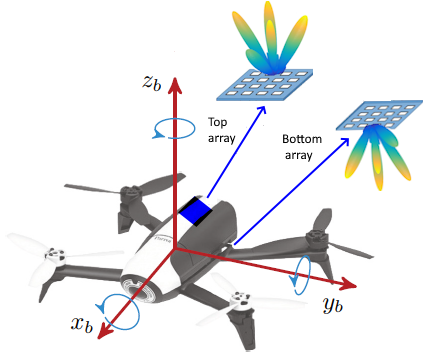}
\hspace{-2em}
\vspace{.19em}
\caption{The body frame $(x_b,y_b,z_b)$ and the 2D antenna arrays on the $x_b-y_b$ plane for a typical UAV. \textcolor{blue}{An omnidirectional antenna is  placed close to each antenna array for side-lobe blocking.}}\label{body_frame}
\end{figure}
\begin{figure}[!t]
\includegraphics[width=3.4in]{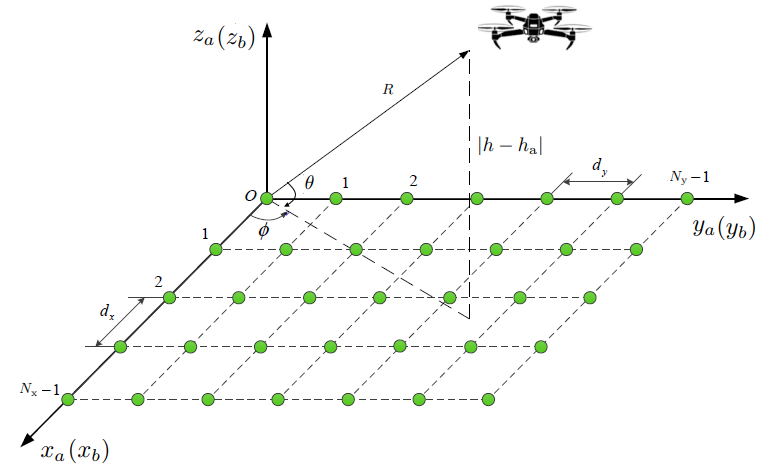}
\caption{The azimuth angle $\phi$ and the elevation angle $\theta$ of the target in the body frame $(x_b,y_b,z_b)$  and the top antenna array frame $(x_a,y_a,z_a)$.}\label{Mode_cccwq89}
\end{figure}
\begin{figure}
  \centering
  \includegraphics[width=8.5cm]{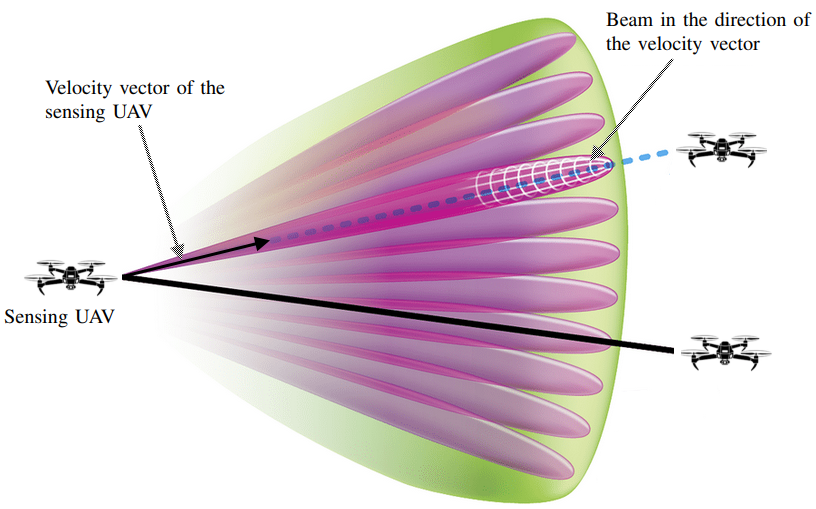}
  \caption{
  The UAV electronically steers the boresight of the antenna array at its top or bottom in the direction of its velocity vector. To increase the situational awareness, the UAV can sequentially scan an extended angular region.
  }\label{2d}
\end{figure}

\begin{figure*}[!t]
\centering
\includegraphics[width=6.5in]{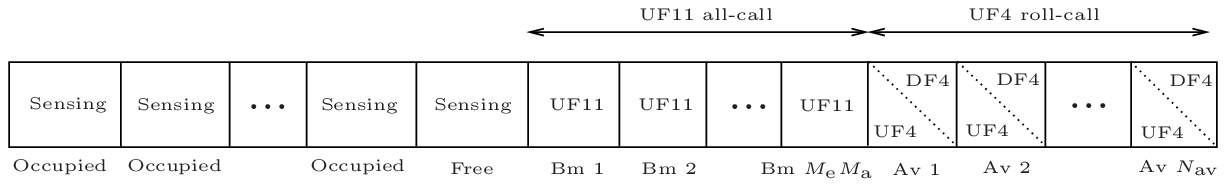}
\caption{The frame structure for RF sensing, the UF11 all-call and roll-call UF4 Mode S interrogations for $M_{\rm e}M_{\rm a}$ beam scans by the UAV. After the reception of the UF4 by the aviation obstacle, it sends the DF4 reply.
Bm and Av stand for beam and aviation obstacle, respectively. The total number of aviation obstacles detected by the UAV during the UF11 all-call is denoted by $N_{\rm av}$.}\label{Mode_frame}
\end{figure*}

\section{System Model}
We consider that UAVs in the airspace are equipped with a miniaturized Mode S transponder, two onboard omnidirectional antennas, and two 
planar antenna arrays with $N_{\rm a}  \triangleq  N_{\rm x} \times N_{\rm y}$ elements, spaced by $d_{\rm x}$ in the rows and $d_{\rm y}$ in the columns.
Hence, a UAV can electronically make a beam scan in azimuth and elevation.
In practice, two antenna arrays are mounted on the top and bottom of the UAV to be able to scan elevation angles in the range of $(-90,90)^{\circ}$ and azimuth angle in the range of $(0,180)^{\circ}$.
An omnidirectional antenna is also placed close to each antenna array for side-lobe blocking.
The UAV switches between the top and bottom antennas according to the direction of its velocity vector. The body frame $(x_b,y_b,z_b)$ and the 2D antenna array planes of a typical UAV are illustrated in Fig. \ref{body_frame}.
As seen, the antenna arrays
are on the $x_b-y_b$ plane.
The onboard antennas enable the transponder to receive the
Mode S interrogations at $1030$ MHz and reply to them at $1090$ MHz.

During navigation, UAVs continuously sense the $1030$ MHz radio frequency band via their omnidirectional antenna. For sensing, a simple energy detector can be adopted.
As soon as the $1030$ MHz channel is identified as unoccupied by a sensing UAV,\footnote{We consider a single channel in this paper; however, by considering multiple channels,  many UAVs can simultaneously initiate Mode S interrogation.}  it initiates the Mode S interrogation by using its antenna array and its omnidirectional antenna \cite{sun20211090}.
The UAV applies electronic beamforming techniques to the antenna array to scan both azimuth angle, $\phi$, and elevation angle, $\theta$, to detect and track the aviation obstacles within the airspace. The azimuth angle, $\phi$, and the elevation angle, $\theta$, of the aviation obstacle in the body frame $(x_b,y_b,z_b)$  and the top antenna array frame $(x_a,y_a,z_a)$ are shown in Fig. \ref{Mode_cccwq89}.

The UAV electronically steers the boresight of the antenna array at its top or bottom in the direction of its velocity vector $(\phi_{\rm v} , \theta_{\rm v} )$. Moreover, in order to increase the situational awareness, it can sequentially scan an extended angular region in the vicinity of the beam in the direction of its velocity vector as shown in Fig. \ref{2d}.
Within the electronically scanned area, the directional antenna boresight (top and bottom)  are located at $(\tilde{\phi},\tilde{\theta})\in {\mathcal S}$ with
\begin{align} \nonumber
{\mathcal S}=&{\Big\{}\big{(}\tilde{\phi}=\phi_{\rm v}+  l\Delta \phi_3,\tilde{\theta}=\theta_{\rm v}+ m\Delta \theta_3 \big{)} \big{|} l,m \in \mathbb{Z},
\\ 
& \tilde{\phi} \in (0,180), \tilde{\theta} \in (-90,90)
\Big{\}},
\end{align}
where $\mathbb{Z}$ is the set of integers, and
$\Delta \phi_3$ and $\Delta \theta_3$ denote the 3dB beamwidth of the array for the azimuth and elevation angles, respectively.

Let $M_{\rm a}$ and $M_{\rm e}$ denote the number of electronic scans sweeping the azimuth and elevation angles, respectively.
 The frame structure for RF sensing, Mode S uplink interrogations, and downlink responses for $M_{\rm e}M_{\rm a}$ beam scans by a UAV are illustrated in Fig.~\ref{Mode_frame}.
As seen, Mode S interrogation is composed of two phases: 1) UF11 all-call, and 2) UF4 roll-call.
In the first phase,
the UAV sequentially scans the airspace with $M_{\rm e}M_{\rm a}$ beams and transmits the UF11 all-call Mode S interrogation within each beam.
During the UF11 all-call interrogation, the sensing UAV obtains the DF11 squitters of all UAVs in its vicinity \cite{sun20211090}.

In the second phase,
after the termination of the UF11 all-call Mode S interrogation for all $M_{\rm e}M_{\rm a}$ beams, the UAV sends the UF4 roll-call Mode S interrogation by selectively addressing the detected aviation obstacles. 
Each aviation obstacle replies to its UF4 roll-call interrogation by broadcasting its altitude information through the DF4 downlink reply from its omnidirectional antenna.
In this interrogation-based scheme, a fast moving aviation obstacle can transmit the DF11 reply in one beam and then leaves the beam. Therefore, it does not receive the UF4 interrogation. This problem can be solved by transmitting the UF4 interrogation in multiple beams regardless of the beamwidth until the UAV receives the DF4 reply of the aviation obstacle.

\begin{figure*}[!t]
\hspace{-1em}
\includegraphics[width=7in]{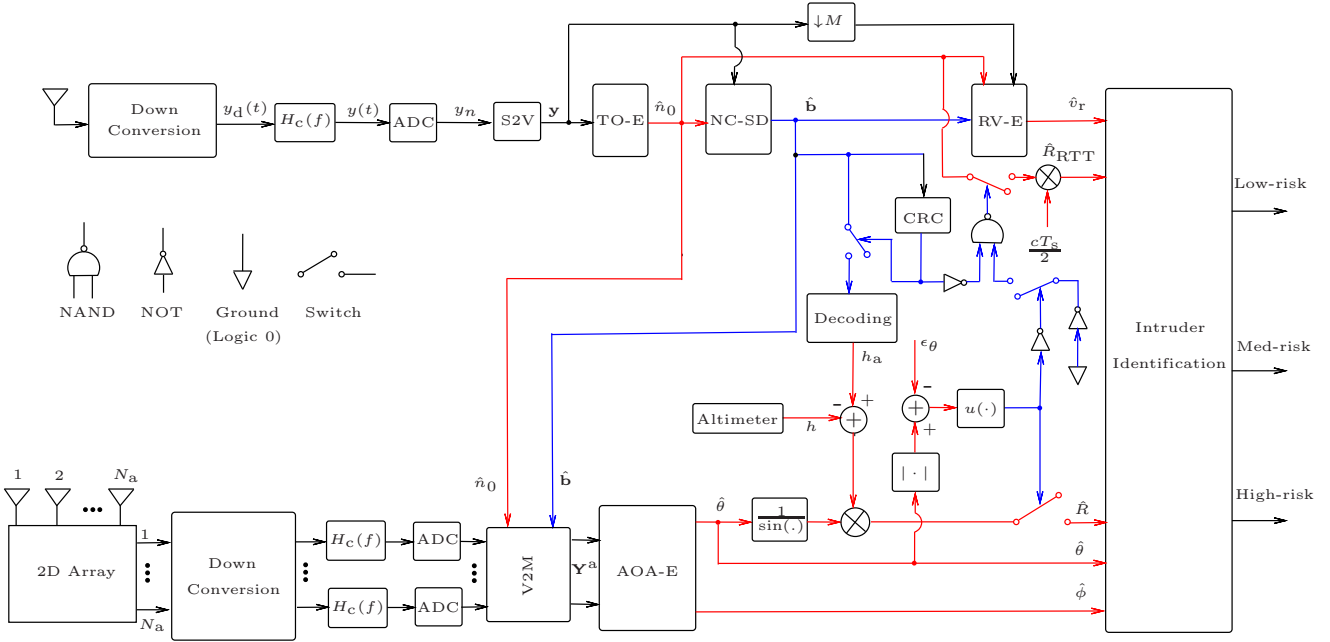}
\caption{Block diagram of the proposed SAA system sensing an aviation obstacle {using its top/bottom antenna array and its top/bottom  omnidirectional antenna.}  The number of antenna elements of the array is $N_{\rm a}=N_{\rm x}N_{\rm y}$, $\downarrow \hspace{-0.3em} M$ denotes the $M$-fold decimation, {$u(\cdot)$ is the unit step function with $u(x)=1$, for $x \geq 0$, and $u(x)=0$, for $x<0$, and $\epsilon_{\theta}$ is a threshold used to switch between the proposed ranging algorithm and the RTT-based ranging.} The abbreviations are: analog-to-digital converter (ADC), Sequence to Vector (S2V), Vector to Matrix (V2M), timing offset estimator (TO-E),
noncoherent symbol detector (NC-SD), radial velocity estimator (RV-E), cyclic redundancy check (CRC), angle-of-arrival estimator (AOA-E). {The black, blue, and red solid lines show the raw data, real values, and logical values, respectively.}}\label{Fig_3787}
\end{figure*}

\section{Proposed Cooperative SAA Method}
We consider perfect RF sensing by the UAVs; hence, only a single UAV emits the UF11 all-call and then the UF4 roll-call Mode S interrogation signals in each navigation step. We propose a cooperative SAA method
that estimates the radial velocity, range, and azimuth and elevation angles of the aviation obstacles from their DF4 roll-call mode S replies. Then, by combining the estimated parameters, the aviation obstacles are classified into high-, medium-, and low-risk intruders for the UAV.
The block diagram of the proposed SAA method sensing an aviation obstacle is illustrated in Fig.~\ref{Fig_3787}.

The TO-E block is the timing offset estimator and is used to estimate the beginning of the received DF4 Mode S reply of the aviation obstacle, i.e., $\hat{n}_0$. The NC-SD block is the noncoherent symbol detector and is employed to demodulate the DF4 reply packet of the aviation obstacle, i.e., $\hat{\bf b}$.
The proposed RV-E block is the radial velocity estimator and is used to estimate the relative radial velocity of the aviation obstacle, i.e., $\hat{v}_{\rm r}$.
The AoA-E block is the angle-of-arrival estimator and is used to estimate the elevation and azimuth angles of the aviation obstacle, i.e., $\hat{\theta}$, and $\hat{\phi}$, respectively.
The decoding block is employed to decode the altitude information of the aviation obstacle, i.e., $h_{\rm a}$, from $\hat{\bf b}$,  and
the cyclic redundancy check (CRC) block is used to validate the integrity of the data in $\hat{\bf b}$. If the CRC verifies the validity of $\hat{\bf b}$,
it is passed through the decoding block to obtain the altitude of the intruder. The elevation angle and the altitude of the aviation obstacle are employed to estimate range $\hat{R}$. If the CRC does not verify the validity of $\hat{\bf b}$ or $|\hat{\theta}|<\epsilon_{\theta}$, the RTT-based ranging is employed.
The outputs of the parameter estimators, i.e.,
$\hat{v}_{\rm r}$,  $\hat{R}$, $\hat{\theta}$, $\hat{\phi}$
are used as inputs of the intruder identification unit.
In the following, the above mentioned blocks and the intruder identification unit are studied in details.
\subsection{Mode S DF4 Discrete Time Signal Model}
Let us consider that the DF4 packet of an aviation obstacle
is received at the UAV omnidirectional antenna with a time delay of $\tau \in  (0, \tau_{\rm max}]$ in
the receiver time reference, where $\tau$ is unknown, and $\tau_{\rm max}$ is known and denotes the maximum delay between the transmission of the
UF4 Mode S interrogation signal by the UAV and the reception of the DF4 reply from an aviation obstacle.

As shown in Fig.~\ref{Fig_3787}, the downconverted DF4 reply signal at the UAV omnidirectional
antenna, i.e.,  $y_{\rm d}(t)$, is passed through
the low-pass filter with frequency response $H_{\rm c}(f)=|H(f)|\exp(j(\angle H(f)-2\pi f\tau_{\rm gr}))$, $f \in [-B,B]$, where $B$ is the bandwidth of the low-pass filter.  The impulse response of the filter is $h(t-\tau_{\rm gr})$, where  $\tau_{\rm gr}-\frac{1}{2\pi}\frac{{\rm d}\angle H(f)}{{\rm d}{f}}$ is the group delay of the filter and $h(t) \triangleq \mathcal{F}^{-1}\big{\{}|H(f)|\exp(j\angle H(f)\big{\}}$ with $\mathcal{F}^{-1}$ as the inverse Fourier transform. 
 Without loss of generality, we assume that $
E_{\rm h} =\int_{-\infty}^{+\infty} h^2(t) {\rm d}t= \int_{-B}^{+B} |H(f)|^2 {\rm d}f =1.
$
The complex baseband filtered signal $y(t)$ can be written as\footnote{{We include the filtering and sampling procedures to emphasize that ideal PPM waveform cannot be considered in the analysis of DF4 reply signal at the receiver.}} \cite{leonardi2017air}
\begin{align}\label{cont}
& y(t)  =  \frac{\sqrt{P_{\rm t}\eta_0}}{L} \Bigg{[} \sum_{m=0}^{15} r_m g\Big{(}t-mT-\frac{T}{2}-\tau_{\rm gr} - \tau\Big{)} \\ \nonumber
&+  \sum_{m=9}^{64} g\Big{(}t-2mT+b_{(m-9)}T+\frac{T}{2}-\tau_{\rm gr}-\tau\Big{)} \Bigg{]} e^{j(2\pi f_{\rm D}t +\psi_0)}
\\ \nonumber
&+  w(t),\,\,\,\,\,\,\,\,\,\,\,\,\,\,\,\,\,\,\,\,\,\,\,\,\,\,\,\,\,\,\,\,\,\,\,\,\,\,\,\, 0 \leq t \leq 128 T+\tau_{\rm gr}+\tau_{\rm max}.
\end{align}
The first term in \eqref{cont} denotes the waveform of the DF4 reply associated with the preamble, and the second term denotes the waveform associated with the pulse-position modulation (PPM) signaling.
For Mode S \cite{sun20211090},
\begin{align}\label{preamble}
{\bf r} &= [r_0 \ r_1 \ \ldots \ r_{15}]^{\rm T} \\ \nonumber
&\triangleq \big{[} 1 \ 0  \ 1 \ 0 \  0 \ 0 \ 0 \ 1 \ 0  \ 1 \ 0 \  0 \ 0 \ 0 \  0 \ 0 \ \big{]}^{\rm T}
\end{align}
which denotes the preamble vector, and
\begin{align}\label{payload}
{\bf b} = [b_0 \ b_1 \ \ldots \ b_{55}]^T
\end{align}
contains the surveillance, communication, and control data along with the parity bits, called payload bits.
In \eqref{cont}, $P_{\rm t}$, $\eta_0$, $2T=1\mu {\rm s}$, $f_{\rm D}$, and $\psi_0$ denote
the peak transmit power, the radiation efficiency of the omnidirectional antenna, the signaling time of the PPM, the Doppler frquency shift, and the carrier phase offset, respectively.
Moreover, $L$ denotes the free space pathloss between the aviation obstacle and the UAV defined as
$
L \triangleq \frac{4\pi R}{\lambda},
$
where $\lambda$ and $R$ denote the carrier wavelength and the distance between the aviation obstacle and the UAV, respectively. The pulse $g(t)$ in \eqref{cont} is given by
$g(t) \triangleq z(t) \circledast h(t)$,
where  $z(t)$ is the Mode S trapezoidal transmit pulse defined as \cite{leonardi2017air}
\begin{align}\label{trap}
 z(t) = \left\{ \begin{array}{l}
\frac{A(t+\frac{T}{2})}{{{\tau _{\rm{r}}}}}\,\,\,\,\,\,\,\,\,\,\,\,\,\,\,\,\,\,\,\,\,\,\,\,\,\,\,\,\ -\frac{T}{2} \le t < -\frac{T}{2} + {\tau _{\rm{r}}},\\
A\,\,\,\,\,\,\,\,\,\,\,\,\,\,\,\,\,\,\,\,\,\,\,\,\,\,\,\,\,\,\,\,\,\,\,\,\,\  -\frac{T}{2}+{\tau _{\rm{r}}} \le t \le \frac{T}{2}- {\tau _{\rm{r}}},\,\,\,\,\,\,\,\,\,\,\,\,\,\,\\
\frac{A( -  t + \frac{T}{2} )}{{{\tau _{\rm{r}}}}}\,\,\,\,\,\,\,\,\,\,\,\,\,\,\,\,\,\,\,\,\,\,\,\,\,\,\,\,\,\,\,\  \frac{T}{2}-\tau_{\rm r},  < t \le \frac{T}{2},   \,\,\,\,\,\,\,\,\,\,\,\,\,
\end{array} \right.
\end{align}
where the rise and decay time of the pulse $z(t)$ is $\tau_{\rm r}$,
and $A$ is a constant value that satisfies
$P_{\rm z} =\frac{1}{T}\int_{-\frac{T}{2}}^{\frac{T}{2}} z^2(t) {\rm d}t=1.$
The average received power at the receiver is given by $P_{\rm av} \approx  \frac{P_{\rm t}\eta_0}{2L^2T} \int_{-T/2}^{T/2} g^2(t) {\rm d}t$,
and the power spectral density (PSD)  of the additive noise $w(t)$ is $N_0$ {watts per hertz}  over the low-pass filter bandwidth, i.e., $f \in [-B,B]$. {In cooperative and RF-based sense and avoid systems, moderately high signal-to-noise (SNR) is required to achieve low bit error rate and thus reliable noncoherent DF4 packet decoding \cite{Agrege}.}

The received signal in \eqref{cont} is sampled with sampling rate $R_{\rm s}=1/T_{\rm s}$, where $T_{\rm s}$ is the sampling time. The discrete received baseband signal for the omnidirectional antenna can be written as
\begin{align}\label{discrete}
{\bf y} \approx \frac{\sqrt{P_{\rm t}\eta_{\rm 0}}}{L} {\bf F}{\bf s}+{\bf w},
\end{align}
where ${\bf F} \triangleq {\rm {diag}}({\bf f})$,
\begin{subequations}
\begin{align}
{\bf f} \triangleq & \big{[}e^{j\psi_0} \ e^{j(2\pi f_{\rm D}T_{\rm s}+\psi_0)}
 \dots \
e^{j(2\pi f_{\rm D}T_{\rm s}(N-1)+\psi_0)}
\big{]}^{\rm T}, \\
{\bf y} \triangleq & \big{[}y_0 \ y_1 \ \ldots \ y_{N-1} \big{]}^{\rm T}, \\
{\bf s} \triangleq & \big{[}{\bf 0}_d^{\rm T} \ ({\bf q} \otimes {\bf g})^{\rm T} \ {\bf 0}_{D-d}^{\rm T}\big{]}^{\rm T}, \\
{\bf q} \triangleq & \big{[}{\bf r}^{\rm T} \ b_0 \ \neg{b}_0 \ b_1 \ \neg{b}_1 \ \ldots \  b_{55} \ \neg{b}_{55} \big{]}^{\rm T}, \\\label{guiio21}
{\bf g} \triangleq & \big{[}g_0 \ g_1 \ \ldots \ g_{M-1} \big{]}^{\rm T}, \\ \label{guiio}
{\bf w} \triangleq & \big{[}w_0 \ w_1 \ \ldots \ w_{N-1} \big{]}^{\rm T},
\end{align}
\end{subequations}
$y_k \triangleq y(kT_{\rm s})$, $w_k \triangleq w(kT_{\rm s})$, $k=0,1,\ldots,N-1$, $g_m \triangleq g(mT_{\rm s}-\frac{T}{2})$, $m=0,1,\dots,M-1$,
and $N \triangleq 128M+D$ with
\begin{align}
M\triangleq \Big{[}\frac{T}{T_{\rm s}}\Big{]}, \,\,\   D = \Big{[}\frac{\tau_{\max}+\tau_{\rm gr}}{T_{\rm s}}\Big{]}+1,  \,\,\ d= \Big{[}\frac{\tau+\tau_{\rm gr}}{T_{\rm s}}\Big{]}+1,
\end{align}
where $2M$ is the number of samples per PPM symbol time $2T$.
For the practical square-root raised cosine low-pass filter $H(f)$ with bandwidth $B$ and roll-off factor $\beta$, the sampling rate $R_{\rm s} = 2B / {(1+\beta)}$ results in independent and identically distributed additive noise at the receiver. In this case,
the noise vector $\bf{w}$ in \eqref{discrete} is zero-mean additive white Gaussian noise (AWGN) with variance $\sigma_{\rm w}^2=N_0E_{\rm h}=N_0$,
where $E_{\rm h}=1$ is the energy of the low-pass filter $h(t)$.
 The values of $\psi_0$ and $f_{\rm D}$ are unknown at the receiver.

\subsection{TO-E and NC-SD Blocks}
In this subsection, we shortly study the TO-E and NC-SD blocks because their outputs are used in the proposed radial velocity estimator and the ranging algorithm as seen in Fig.~\ref{Fig_3787}.

{\it{TO-E}:}
For exclusive SAA applications,  because the response of the aviation obstacle to the UF4 roll-call Mode S interrogation is DF4 reply and this is known to the UAV, the preamble vector in \eqref{preamble} along with the 5 PPM symbols denoting the format number of the packet can be employed for time synchronization; thus, the preamble vector used for time synchronization can be extended to
\begin{align}\label{ex_preamble}
\bar{\bf r}\triangleq
\big{[}\bar{r}_0 \ \bar{r}_1 \ \ldots \ \bar{r}_{25} \big{]}^{\rm T} = \big{[} {\bf r}^{\rm T} \ 0 \ 1 \ 0 \ 1 \ 1 \ 0 \ 0 \ 1 \ 0 \ 1 \big{]}^{\rm T},
\end{align}
where ${\bf r}$ is given in \eqref{preamble}.
By employing the correlation estimator, the beginning of the DF4 packet $n_0$ can be straightforwardly estimated as \cite{sun20211090}
\begin{equation}\label{to_estimate}
\begin{array}{rrclcl}
\hat{n}_0 = \displaystyle
 \operatorname*{argmax}_{n_0} & \Big{|}\sum_{n=n_0}^{n_0+25}y_n^* \bar{r}_{n-n_0}
\Big{|}^2,
\\
 \textrm{s.t.} &  M_{\rm gr} \leq n_0 \leq D\\
\end{array}
\end{equation}
where $M_{\rm gr} \triangleq [ \frac{\tau_{\rm gr}}{T_{\rm s}}]+1$.

{\it{NC-SD}:}
After estimating the beginning of the DF4 response packet, the PPM data symbols can be noncoherently detected as
\begin{align}\label{decode_ppm}
\hat{b}_n=
\left\{
	\begin{array}{ll}
		1  & \mbox{if } \sum_{m=0}^{M-1} \Lambda_{n,m} \geq \sum_{m=M}^{2M-1}\Lambda_{n,m}  \\ \\
		0 & \mbox{if } \sum_{m=0}^{M-1} \Lambda_{n,m} < \sum_{m=M}^{2M-1}\Lambda_{n,m},
	\end{array}
\right.
\end{align}
for $n=0,1,\ldots,55$, where $\hat{b}_n$ is the $n$th decoded data bit, and
$\Lambda_{n,m} \triangleq \big{|}y_{\hat{n}_{\rm s}+2nM+m}\big{|}^2$, $\hat{n}_{\rm s} \triangleq \hat{n}_0+16M$.
By decoding the detected bits $\hat{b}_n$ in \eqref{decode_ppm}, the altitude of the aviation obstacle is obtained. The altitude information can be used for radial velocity estimation and ranging.

\begin{figure}[!t]
\includegraphics[width=3.8in]{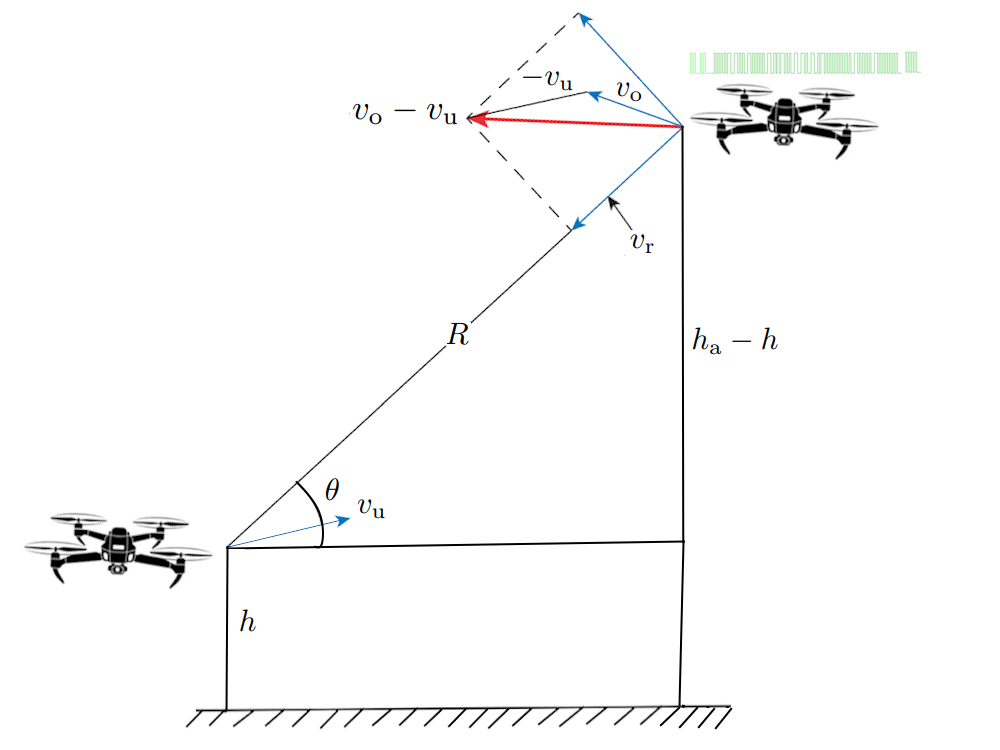}
\caption{{The velocity vector of the UAV, $v_{\rm u}$, the velocity vector of the aviation obstacle, $v_{\rm o}$,
the relative velocity vector $v_{\rm o}-v_{\rm u}$, the radial velocity vector $v_{\rm r}$, the altitude of the UAV, $h$, and the altitude of the aviation obstacle, $h_{\rm a}$. }}\label{Mode_cccwq8u9}
\end{figure}

\section{Proposed Radial Velocity Estimator}
It this section, we investigate in detail the RV-E block in Fig.~\ref{Fig_3787} and  show  how the detected DF4 reply can be employed for radial velocity estimation of the aviation obstacles. The proposed radial velocity estimator is one of the main contributions of this paper.

The Doppler frequency shift $f_{\rm D}$ in \eqref{cont} can be expressed as a function of the relative radial velocity of the aviation obstacle and UAV, $v_{\rm r}$, as 
\begin{align}\label{radial_velocity}
f_{\rm D}=\frac{v_{\rm r}f_{\rm c}}{c},
\end{align}
where $c = 3 \times {10}^8$ m/s and $f_{\rm c}=1090 \times 10^{6}$ Hz denote the speed of light and the carrier frequency of the DF4 reply, respectively.
{The relative radial velocity of the UAV and an aviation obstacle is shown in Fig.~\ref{Mode_cccwq8u9}.}
For the derivation of the radial velocity estimator, we consider perfect time delay estimation, i.e., $\hat{n}_0=n_0$, and perfect data detection, i.e.,
$\hat{\bf b}={\bf b}$.
However, in practice, the estimated time delay and the detected PPM data are replaced in the derived radial velocity estimator.

Let $n_0$ denote the index of the first sample of the received DF4 reply. The index of the sample denoting the start of the payload symbols is $n_{\rm s} \triangleq n_0+16M$ because the length of the sampled preamble vector in \eqref{preamble} is $16M$. Let us write the PPM form of the payload vector $\bf{b}$ in \eqref{payload} as
\begin{align}\label{ppm_bits}
\bar{\bf b} &= \big{[}\bar{b}_0 \ \bar{b}_1 \ \ldots \  \bar{b}_{111}  \big{]}^{\rm T} \\ \nonumber
& \triangleq \big{[}b_0 \ \neg{b}_0 \ b_1 \ \neg{b}_1 \ \ldots \  b_{55} \ \neg{b}_{55} \big{]}^{\rm T},
\end{align}
and let ${\mathcal I}$ denote the set that includes the index of the non-zero elements of the vector $\bar{\bf b}$, defined as
\begin{align}\label{set}
{\mathcal I} = \{i_0,i_1,\ldots,i_{i_{55}}\}
\triangleq \{i \in \{0,1,\ldots,111\}|\bar{b}_i \neq 0\}.
\end{align}
The cardinality of ${\mathcal I}$ is $|{\mathcal I}|=56$ because of the PPM signaling of $\bar{\bf b}$.
In Theorem \ref{doppler}, we propose a low-complexity and accurate Doppler frequency shift estimator using the DF4 reply of an aviation obstacle.
\begin{theorem}\hspace{-0.7em}:\label{doppler}
At moderately high SNR values
that is the case in cooperative and RF-based SAA systems, the Doppler frequency shift ${f}_{\rm D}$
between the UAV and an aviation obstacle can be estimated by
\begin{align}\label{estimator}
 \hat{f}_{\rm D}  = \frac{\hat{v}_{\rm r} f_{\rm c}}{c} =\frac{1}{2\pi M T_{\rm s}} \frac{{\bf u}^{\rm T} {\bm \Delta}^{\rm{T}} {\bf C}^{-1}{\bm \Delta} {\bm \varphi}}{{\bf u}^{\rm T}{\bm \Delta}^{\rm T} {\bf C}^{-1}{\bm \Delta}{\bf u}},
\end{align}
where
\begin{subequations}
\begin{align}
{\bm u} &\triangleq \big{[} {i_0}  \ {i_1} \ \ldots \ {i_{55}}\big{]}^{\rm T},
\\
{\bm \varphi} &\triangleq \big{[} \measuredangle x_{i_0}  \ \measuredangle x_{i_1} \ \ldots \ \measuredangle x_{i_{55}}\big{]}^{\rm T},
\\
x_i & \triangleq y_{n_{\rm d}+iM},
\\
n_{\rm d} & \triangleq n_0+16M+[M / 2],
\end{align}
\end{subequations}
and the elements of the vector ${\bm u}$, are given in \eqref{set} with $i_0<i_1<\ldots<i_{55}$. In \eqref{estimator}, ${\bm \Delta }$ is the $55\times 56$ discrete derivative matrix as
\begin{align}\label{derivative}
{\bm \Delta } \triangleq
\left[ {\begin{array}{*{20}{c}}
1&{ - 1}&0&0&0& \cdots &0\\
0&1&{ - 1}&0&0& \cdots &0\\
 \vdots & \ddots & \ddots & \ddots & \ddots & \cdots & \vdots \\
0&0& \cdots &1&{ - 1}&0&0\\
0&0& \cdots &0&1&{ - 1}&0\\
0&0& \cdots &0&0&1&{ - 1}
\end{array}} \right],
\end{align}
and ${\bf C}$ is the $55\times 55$ sqaure matrix as
\begin{align}
{\bf C}=\sigma_{\rm \vartheta}^2
\left[ {\begin{array}{*{20}{c}}
2&{ - 1}&0&0\\
{ - 1}&2&{ - 1}&0\\
 \cdots & \cdots & \cdots & \cdots \\
0&0& \cdots &0
\end{array}\,\,\,\,\,\begin{array}{*{20}{c}}
 \cdots \\
 \cdots \\
 \cdots \\
{ - 1}
\end{array}\,\,\,\,\,\begin{array}{*{20}{c}}
0\\
0\\
 \cdots \\
2
\end{array}} \right],
\end{align}
with
\begin{align}
\sigma_{\rm \vartheta}^2 \triangleq \frac{1}{2} \Big{(}\frac{L \sigma_{\rm w}}{\sqrt{P_{\rm t}\eta_{\rm 0}g_{n_{\rm c}}}}\Big{)}^2,
\end{align}
and $g_{n_{\rm c}}$ with $n_{\rm c} \triangleq [M / 2]$, is the $(n_{\rm c}+1)$th elements of the pulse shaping vector ${\bf g}$ given in \eqref{guiio21}.

The Doppler frequency shift estimator in \eqref{estimator} is an unbiased estimator, i.e., $\mathbb{E}\{\hat{f}_D\}=f_{\rm D}$, and its RMSE is given by
\begin{align}\label{var}
\sqrt{\mathbb{E}\{(f_{\rm D}-\hat{f}_{\rm D})^2\}} = \sqrt{\frac{1}{(2\pi MT_{\rm s})^2} \frac{1}{{\bf u}^{\rm T}{\bm \Delta}^{\rm T}{\bf C}^{-1}{\bm \Delta}{\bf u}}}.
\end{align}
\end{theorem}
\hspace{21em} $\blacksquare$

{\it Proof in Appendix \ref{Appx_A}}.

Note that the inverse of the matrix ${\bf C}$ is given by
$
[{\bf C}^{-1}]_{mn}=\sigma_{\rm \vartheta}^{-2} \big{[}\min(m,n)-\frac{mn}{56}\big{]},
$
where $1 \leq m,n \leq 55$ and $\min(m,n)$ denotes the minimum of $m$ and $n$ \cite{marcus1948basic}. As seen, the estimator in \eqref{estimator} does not require a priori knowledge on $\sigma_{\rm \vartheta}^2$ and thus the path loss $L$ and the transmit power $P_{\rm t}$.

\begin{prop}\hspace{-0.5em}:
{Maximum Unambiguous Radial Velocity}
\end{prop}
PPM signaling can be considered as waveform of pulse radar with random pulse repetition intervals (PRI)s of length  $T$, $2T$,
and $3T$,  and pulse width $T$.  
Hence, the maximum unambiguous Doppler frequency shift  that can be estimated is given by 
\begin{align}\label{amb-rang}
|f_{\rm D}| \leq \frac{1}{T},
\end{align}
which results in the maximum relative radial velocity of 
\begin{align}\label{xxy}
|v_{\rm r}| \leq \frac{\lambda}{T}.
\end{align}
{\it Proof in Appendix \ref{Appx_C}}.

\begin{prop}\hspace{-0.5em}:
 {Separation of the Doppler Frequency Shift and the Frequency Drift}
\end{prop}
In addition to Doppler frequency shift due to the relative radial velocity between the UAV and the aviation obstacle, there is always some difference between the manufacturer’s nominal operational frequency and the real frequency of the Mode S transponder. Moreover, the actual frequency keeps slightly changing with temperature, pressure, age, and some other factors. Hence, a
frequency drift between the UAV and the aviation obstacle oscillators is expected.

Distinction between the frequency drift and the Doppler frequency shift is impossible for a single-carrier communication, as it is currently in Mode S SSR.
However, the frequency drift and the Doppler frequency shift can be separated in multicarrier communication with at least two subcarriers.
Hence, for applications that high precision radial velocity estimation is needed, the DF4 replies are needed to be transmitted in the form of multicarrier.

Let $df$ denote the frequency drift between the transmitter and receiver oscillators of a multicarrier system with subcarrier frequencies
$f_k \triangleq  Q_k\Delta f$, $k \in \{0, 1, \ldots , {N_{\rm c}-1}\}$ in the baseband, where $\Delta f$ and $N_{\rm c}$, denote the subcarrier bandwidth and the number of subcarriers, respectively.
Because of the frequency drift, the frequency offset of the $k$th subcarrier ${f}_k$  can be written as
\begin{align}\label{yuoptmt}
{f}_k & \approx \frac{v_{\rm r} (f_{\rm c} + Q_k\Delta f )}{c} + df,
\end{align}
where $k=0,1,\ldots,N_{\rm c}-1$,  $f_{\rm c}$ is the carrier frequency, 
and $c$ denotes the speed of light. 

Let $\hat{f}_k$ denote the estimation of $f_k$ by using the proposed estimator in \eqref{estimator}.
We can write $\bm{\Omega} \triangleq [\hat{f}_0 \ \hat{f}_1 \ \ldots \ \hat{f}_{N_{\rm c}-1}]^{\rm T}$ in vector form as
\begin{align}
\bm{\Omega} = {\bm \Psi}{\bm \eta}+{\bf z}
\end{align}
where ${\bf z}$ is the additive noise vector, ${\bm \eta} \triangleq [df \ v_{\rm r}]^{\rm T}$,  and
\begin{align}
{\bm \Psi} \triangleq
\left[{\begin{array}{*{20}{c}}
1&{\frac{{{f_{\rm{c}}} + {Q_0}\Delta f}}{c}}\\
1&{\frac{{{f_{\rm{c}}} + {Q_1}\Delta f}}{c}}\\
 \vdots & \vdots \\
1&{\frac{{{f_{\rm{c}}} + {Q_{{N_{\rm{c}}} - 1}}\Delta f}}{c}}
\end{array}} \right].
\end{align}
The LS estimate of $\bm{\eta}$ is expressed as
\begin{equation}
\begin{array}{rrclcl}
\hat{{\bm \eta}} = \displaystyle
 \operatorname*{argmax}_{\bm \eta} & \big{\|}\bm{\Omega} - {\bm \Psi}{\bm \eta}\big{\|}^2 ,
\end{array}
\end{equation}
which yields $\hat{{\bm \eta}}= [\hat{df} \ \hat{v}_{\rm r}]^{\rm T}= ({\bm \Psi}^{\rm T}{\bm \Psi})^{-1}{\bm \Psi}^{\rm T}\bm{\Omega}$.
It is obvious that the Doppler frequency shift of the $k$th subcarrier is estimated as ${\hat{v}_{\rm r} (f_{\rm c} + Q_k\Delta f )}/{c}$.

\section{AOA Estimation}
In this section, we investigate the AOA-E block in Fig.~\ref{Fig_3787}. The estimated elevation angle of the aviation obstacle $\theta$ is used in the proposed ranging algorithm in Section  \ref{ranging}. To estimate the azimuth and elevation angles of aviation obstacles, the SAA system uses its 2D planar antenna array.
For AOA estimation, we use the $56M$ received samples corresponding to the elements of $\bar{\bf b}$ in \eqref{ppm_bits} with the indices in the set $\mathcal{I}$ in \eqref{set}. 
In this case,
the received discrete baseband matrix in the antenna array with $N_{\rm a} \triangleq  N_{\rm x}N_{\rm y}$ antenna elements can be written as
\begin{align}\label{array}
{\bf Y}^{\rm a} &= \big{[} {\bf y}_1^{\rm a} \ {\bf y}_2^{\rm a} \  \ldots {\bf y}_{N_{\rm a}}^{\rm a} \big{]} \\ \nonumber
& \approx
\frac{\alpha f(\theta,\phi) \sqrt{P_{\rm t}\eta}}{L}{\bf F}^{\rm a}{({\bf 1}_{56}\otimes {\bf g})}{\bf a}^{\rm T}(\theta,\phi)+{\bf Z}^{\rm a},
\end{align}
where ${\bf Y}^{\rm a}$ is the $56M \times N_{\rm a}$ measurement matrix, $f(\theta,\phi)$ is the normalized field pattern\footnote{For antenna array with omnidirection elements, we have $f(\theta,\phi)=1$.} (normalized to unity), $\alpha$ is a nominal field amplitude by the aviation obstacle, $P_{\rm t}$ is the peak transmit power by the aviation obstacle,
$\eta$ is the radiation efficiency of the array,
${\bf F}^{\rm a} \triangleq {\rm diag}({\bf f}^{\rm a})$ with
\begin{align}
{\bf f}^{\rm a} \triangleq & \big{[}{\bf f}_{i_0}^{\rm T} \ {\bf f}_{i_1}^{\rm T}
 \dots \
{\bf f}_{i_{55}}^{\rm T}
\big{]}^{\rm T},
\end{align}
\begin{align} \nonumber
{\bf f}_i \triangleq & \big{[}e^{j(2\pi f_{\rm D}T_{\rm s}(n_{\rm s}+iM)+\psi_{\rm a})} \ e^{j(2\pi f_{\rm D}T_{\rm s}(n_{\rm s}+iM+1)+\psi_{\rm a})}
\\
& \,\,\,\,\,\,\,\,\,\,\,\,\ \dots \
e^{j(2\pi f_{\rm D}T_{\rm s}(n_{\rm s}+(i+1)M-1)+\psi_{\rm a})}
\big{]}^{\rm T},
\end{align}
$n_{\rm s}=n_0+16M$, $i_k \in \mathcal{I}$, the carrier phase offset of the array is $\psi_{\rm a}$,\footnote{Because a single RF oscillator is employed at the antenna array, the carrier phase $\psi_{\rm a}$ is the same for all array elements.}  and ${\bf{1}}_{56}$ denotes all-ones vector of length $56$.  The array manifold vector of the 2D planar antenna array with $N_{\rm x} \times N_{\rm y}$ elements, spaced by $d_{\rm x}$ in the rows and $d_{\rm y}$ in the columns is
$
{\bf a}(\theta,\phi)\triangleq [a_{1,1}(\theta,\phi),a_{2,1}(\theta,\phi),\ldots,a_{N_{\rm x},N_{\rm y}}(\theta,\phi)]^{\rm T}$  \cite{van2002optimum},
where
\begin{align}\label{manifold_op}
a_{m,n}({\theta} & ,\phi)
\\ \nonumber
  & = e^{j\frac{2\pi}{\lambda}(d_{\rm x}(m-1)\cos(\theta)+d_{\rm y}(n-1)\sin(\theta)\sin(\phi))}.
\end{align}
The complex-valued AWGN matrix ${\bf Z}^{\rm a}$ with dimension $56M \times N_{\rm a}$ in \eqref{array} is given by
$
{\bf Z}^{\rm a} \triangleq \big{[} {\bf z}_1^{\rm a} \ {\bf z}_2^{\rm a} \  \ldots \ {\bf z}_{N_{\rm a}}^{\rm a} \big{]},
$
where
$
\mathbb{E}\big{\{}{\bf z}_k^{\rm a} ({\bf z}_\ell^{\rm a})^{\rm H}\big{\}}=\delta[k-\ell]\sigma_{\rm w}^2{\bf I}_{56M},
$
$k,\ell \in \{1,2,  $ $\ldots,N_{\rm a}\}$, and $\sigma_{\rm w}^2 =N_0E_{\rm h}=N_0$.

\subsection{2D-MUltiple SIgnal Classification (2D-MUSIC) Algorithm}
To estimate $\theta$ and $\phi$, different algorithms can be employed. If $P_{\rm t}$ or $R$ are unknown,
the 2D-MUSIC algorithm is an effective method for estimating AOA.
By employing the 2D-MUSIC algorithm, the elevation and azimuth angles of the aviation obstacle can be estimated from its DF4 reply as
\begin{equation}\label{optimization}
\begin{array}{rrclcl}
\{\hat{{\theta}},\hat{{\phi}}\} = \displaystyle
\operatorname*{argmax} & P_{\rm 2D-MUSIC}(\theta,\phi),&\,
\\
\hspace{-8em} \textrm{s.t.}   &  \hspace{-8em} \theta \in [\theta_{\rm l}^{\rm 11},\theta_{\rm u}^{\rm 11}]  \\
                  & \phi \in [\phi_{\rm l}^{\rm 11},\phi_{\rm u}^{\rm 11}]
\end{array}
\end{equation}
where
\begin{align}\label{optimization2}
P_{\rm 2D-MUSIC}(\theta,\phi) \triangleq \frac{1}{\sum_{n=2}^{N_{\rm a}}\big{|}{\bf a}^{\rm H}(\theta,\phi)\hat{\bf{e}}_n\big{|}^2}.
\end{align}
The constraint of the maximization in \eqref{optimization}, i.e., $\phi \in [\phi_{\rm l}^{\rm 11},\phi_{\rm u}^{\rm 11}]$ and $\theta \in [\theta_{\rm l}^{\rm 11},\theta_{\rm u}^{\rm 11}]$, is the angular scanning area, in which
 the UAV sends the all-call UF11 Mode S interrogation signal to the aviation obstacles, and  $\hat{{\bf{e}}}_n$ is the eigenvector corresponding to the $n$th largest eigenvalue of the sample correlation matrix  given by {$\hat{\bf A}=\frac{1}{56M}({\bf Y}^{\rm a})^{\rm H}{\bf Y}^{\rm a}$, where ${\bf Y}^{\rm a}$ is the matrix of received samples given in  \eqref{array}.}
In Appendix \ref{Appx:d}, the performance analysis of the 2D-MUSIC algorithm for the DF4 reply is provided.

\begin{figure*}
\centering
\begin{tabular}{ |p{0.8cm}||p{1.7cm}|p{2.5cm}|p{2.4cm}|p{1.8cm}|}
\hline
 \multicolumn{5}{|c|}{TABLE II: Intruder identification} \\
 \hline
 Class& $({R}<\gamma_{\rm R})$ &$(|{\phi}-{\phi}_{\rm v}|<\gamma_{\phi})$&$(|{\theta}-{\theta}_{\rm v}|<\gamma_{\theta})$& $({v}_{\rm r}>\gamma_{v})$\\
 \hline
 $H_0$   & \,\,\,\,\,\,\,\,\,\,\ 1    &\,\,\,\,\,\,\,\,\,\,\,\,\,\,\,\,\,\ X&  \,\,\,\,\,\,\,\,\,\,\,\,\,\,\,\,\,\  X & \,\,\,\,\,\,\,\,\,\,\ X\\
 $H_0$   & \,\,\,\,\,\,\,\,\,\,\ 0    &\,\,\,\,\,\,\,\,\,\,\,\,\,\,\,\,\,\ 1&  \,\,\,\,\,\,\,\,\,\,\,\,\,\,\,\,\,\,\  1 & \,\,\,\,\,\,\,\,\,\,\,\ 1\\
 $H_1$   & \,\,\,\,\,\,\,\,\,\,\ 0    &\,\,\,\,\,\,\,\,\,\,\,\,\,\,\,\,\,\ 1&  \,\,\,\,\,\,\,\,\,\,\,\,\,\,\,\,\,\,\  0 & \,\,\,\,\,\,\,\,\,\,\,\ 1\\
 $H_1$   & \,\,\,\,\,\,\,\,\,\,\ 0    &\,\,\,\,\,\,\,\,\,\,\,\,\,\,\,\,\,\ 0&  \,\,\,\,\,\,\,\,\,\,\,\,\,\,\,\,\,\,\  1 & \,\,\,\,\,\,\,\,\,\,\,\ 1\\
 $H_2$   & \,\,\,\,\,\,\,\,\,\,\ 0    &\,\,\,\,\,\,\,\,\,\,\,\,\,\,\,\,\ X&  \,\,\,\,\,\,\,\,\,\,\,\,\,\,\,\,\,\  X & \,\,\,\,\,\,\,\,\,\,\,\ 0\\
 \hline
\end{tabular}
\end{figure*}

\section{Proposed Ranging algorithm based on Triangular Relation}\label{ranging}
In this section, we develop a ranging algorithm by using the triangular relation between the elevation angle $\theta$  of the aviation obstacle and its altitude difference with the UAV, $h_{\rm a}-h$. The proposed ranging algorithm is one of the main contributions of this paper.
By employing $\hat{\theta}$ via either the top or the bottom antennas and the altitude information of the aviation obstacle, i.e., $h_{\rm a}$, obtained by DF4 reply decoding, the UAV can estimate the range of the aviation obstacle as
\begin{align}\label{range}
\hat{R}= \frac{h_{\rm a}-{{h}}}{\sin({\hat{\theta}})},
\end{align}
where $h$ is the altitude of the UAV measured by its altimeter.
{The necessary condition for ranging on the basis of triangular relation in \eqref{range} is that the CRC is satisfied and $|\hat{\theta}| >\epsilon_{\theta}$, where 
$\epsilon_{\theta}$ is a predefined threshold.
Otherwise, the SAA system can employ the conventional RTT-based ranging algorithm as shown in Fig.~\ref{Fig_3787}.} It should be noted that for the top (bottom) antenna array, we have $\hat{\theta}>0$ ($\hat{\theta}<0$).
In Theorem \ref{theorem}, we obtain approximate expressions for the mean and RMSE of the range estimator in \eqref{range}.

\begin{theorem}\label{theorem}{\hspace{-1em} :}
At moderately high SNR values
that is the case in cooperative and RF-based SAA systems, the mean and  RMSE of the range estimator in \eqref{range} for $\theta \in (-90,90)$ and $\theta \neq 0$ is given by
\begin{align}
{\mathbb{E} \{\hat{R} \}} = \mathbb{E}\Bigg{\{}\frac{h_{\rm a}-h}{\sin(\hat{\theta})}\Bigg{\}}  \approx \frac{h_{\rm a}-h}{\sin({\theta})}.
\end{align}
and
\begin{align}\label{var_range}
\sqrt{{\mathbb{E} \{(\hat{R}-R)^2\}}} &  \approx \frac{|h_{\rm a}-h|}{\sin^2({\theta})} \\ \nonumber
& \times  \sqrt{\frac{\frac{\sigma_{\rm w}^2}{112M}\frac{\lambda_{\rm max}}{(\sigma_{\rm w}^2-\lambda_{\rm max})^2}\big{|}{\bf a}^{\rm H}(\theta,\phi){\bf{e}}_1\big{|}^2}{\sum_{n=2}^{N_{\rm a}}\big{|}{\bm \alpha}^{\rm H}({\theta},\phi){\bf e}_n\big{|}^2}}.
\end{align}
where ${\bm \alpha}({\theta},\phi)\triangleq \big{[}\alpha_{1,1}({\theta},\phi) \ \alpha_{2,1}({\theta},\phi) \ \ldots \ \alpha_{N_{\rm x},N_{\rm y}}({\theta},\phi)\big{]}^{\rm T}$  with $\alpha_{m,n}({\theta},\phi)$ as
\begin{align}\label{qqqty}
&\alpha_{m,n}({\theta},\phi) \triangleq \frac{\partial a_{m,n}({\theta},\phi)}{\partial \sin(\theta)}\\ \nonumber
& =  \frac{j2\pi}{\lambda}\Big{(} d_{\rm y}(n-1)\sin(\phi)-\tan(\theta)d_{\rm x}(m-1) \Big{)}a_{m,n}({\theta},\phi),
\end{align}
and $a_{m,n}({\theta},\phi)$ is given in \eqref{manifold_op}. In \eqref{var_range},
the vector ${{\bf{e}}}_1$ is the eigenvector corresponding to the largest eigenvalue of ${\bf \Lambda} \triangleq \xi{\bf a}^*(\theta,\phi){\bf a}^{\rm T}(\theta,\phi)+\sigma_{\rm w}^2{\bf I}$, i.e., $\lambda_{\rm max}$,  ${{\bf{e}}}_k$, $k=2,3,\ldots,N_{\rm x}N_{\rm y}$ are the eigenvectors corresponding to the $N_{\rm x}N_{\rm y}-1$ eigenvalues $\lambda=\sigma_{\rm w}^2$, and
\begin{align}\label{eq:p}
\xi \triangleq \frac{\alpha^2 f^2(\theta,\phi)P_{\rm t}\eta \sum_{k=1}^{M}g_k^2}{L^2M}
\\ \nonumber
\approx \frac{\alpha^2 f^2(\theta,\phi)P_{\rm t}\eta\int_{-\infty}^{\infty}g^2(t){\rm d}t}{T L^2}.
\end{align}
The approximation in \eqref{eq:p} is obtained for large enough $M$, where  $\frac{1}{M} \sum_{k=1}^{M}
g_k^2\approx$ $ \frac{1}{T}\int_{-\infty}^{\infty}g^2(t){\rm d}t$.
\end{theorem}
\hspace{21em} $\blacksquare$

{\it
Proof in Appendix \ref{Appx_B}}.

\section{Intruder Identification}
Let ${\phi}_{\rm v}$ and ${\theta}_{\rm v}$ denote the direction of the UAV velocity vector in the antenna array coordinate.
By estimating $\hat{v}_{\rm r}$, $\hat{\phi}$, $\hat{\theta}$, and $\hat{R}$  through \eqref{estimator}, \eqref{optimization}, and \eqref{range},
the UAV obtains the necessary information for classification of the aviation obstacles.
The typical criteria used to evaluate the risk of a collision depend
on the class of UAVs.
In this paper, we propose an intruder identification mechanism in which the UAV uses $\hat{v}_{\rm r}$, $\hat{\phi}$, $\hat{\theta}$, and $\hat{R}$ to classify aviation obstacles into three classes: 1) high-risk intruder $H_0$, 2) medium-risk intruder $H_1$, 3) and low-risk intruder $H_2$. We assume that the angular velocities of the aviation obstacle in the antenna array coordinate, i.e., $|{\rm d}\phi / {\rm d}t|$ and $|{\rm d}\theta / {\rm d}t|$ are small.

In SAA systems, a safety radius $\lambda_R$ is considered for safe navigation,
and if the distance between the
UAV and the aviation obstacle is smaller than this safety radius,
the risk of collision is significantly high regardless of radial velocity $v_{\rm r}$, the relative azimuth angle $|\phi-\phi_{\rm v}|$, and relative elevation angle $|\theta-\theta_{\rm v}|$ \cite{woo2020collision}. The aviation obstacle within the safety radius is identified as a high-risk intruder. Furthermore,
in the case that
the aviation obstacle is outside the safety radius, the radial velocity, $v_{\rm r}$, exceeds a safety threshold $\gamma_{\rm v}$,  and the UAV and the aviation obstacle are moving towards each other, i.e., $(|{\phi}-{\phi}_{\rm v}|<\gamma_{\phi})$ and $(|{\theta}-{\theta}_{\rm v}|<\gamma_{\theta})$, the aviation obstacle is also identified as a high-risk intrusion because the risk of collision is still high.

In the case that the radial velocity, $v_{\rm r}$,  exceeds the safety threshold, $\gamma_{\rm v}$,  and the UAV and aviation obstacle are not moving towards each other, i.e., $(|{\phi}-{\phi}_{\rm v}|>\gamma_{\phi})$ or $(|{\theta}-{\theta}_{\rm v}|>\gamma_{\theta})$, the aviation obstacle is identified as a medium-risk intruder because it has the potential to rapidly reach the safety radius. Finally, in the case that  the aviation obstacle is outside the safety radius, and the radial velocity, $v_{\rm r}$, does not exceed the safety threshold $\gamma_{\rm v}$, the aviation obstacle is identified as low-risk intruder.

Given the parameters ${\phi},{\theta},{R},$ and ${v}_{\rm r}$, the proposed intruder identifier is given  in Table II,
where $\gamma_R$, $\gamma_{\phi}$, $\gamma_{\theta}$, and $\gamma_v$ are the predefined safety thresholds. The value of these thresholds varies
for different classes of UAVs. 
In Table II, $1$ and $0$ are used to represent the true and false conditions, respectively, and $X \in \{0,1\}$. During the SAA procedure,  the estimated parameters $\hat{\phi},\hat{\theta},\hat{R},$ and $\hat{v}_{\rm r}$ are substituted in the classification Table II to identify the risk level of the aviation obstacle.
The summary of the proposed SAA method for $N_{\rm av}$ detected aviation obstacles is summarized in Algorithm \ref{algorithmrt}.

   \begin{table}
   
    \caption{{Simulation setup parameters}} 
    \centering 

    \begin{tabular}{| c | c | c | c |} 
    \hline\hline 
    Parameter & Unit & Value \\
    [0.5ex]
    \hline 
    $P_{\rm t}$ & dBW & 8 \\
    $B$ & MHz & 6 \\
    $T$ & $\mu$s & 0.5 \\
    $\tau_{\rm r}$ & $\mu$s & $0.01$ \\
    $\tau_{\rm gr}$ & $\mu$s & $270$ \\
    $\tau_{\rm max}$ & ms & $21.4$ \\
    $\Delta \phi_3$  & Degree & 60 \\
    $\Delta \theta_3$  & Degree & 60  \\
    $N_0$  & watts/Hz& $2.4\times 10^{-21}$ \\
    $d_{\rm x}$ & m & $0.1375$\\
    $d_{\rm y}$ & m & $0.1375$\\
    $N_{\rm a}$ & - & $4$\\
    $N_{\rm x}$ & - & $2$\\
    $N_{\rm y}$ & - & $2$\\
    $M$ & - & $3$\\
    $\epsilon_{\theta}$ & Degree & $1$\\
    \hline 
    \end{tabular}
    \label{tab:orbital_data}
    \end{table}

\begin{algorithm}
  \KwResult{Risk status and navigation parameters of the aviation obstacles, i.e., $\mathcal{R}$ and  $\mathcal{P}$}
  Initialization:\  \text{$\mathcal{R}=\{ \}$ and  $\mathcal{P}=\{ \}$}\;
  Sense the $1030$ MHz radio frequency band using omnidirectional antenna\;
  \eIf{ Channel is not occupied}{
  \For{$m=1,2,\ldots,M_{\rm e}M_{\rm a}$}{
        Transmit UF11 Mode S all-call interrogation for the $m$th beam space\;
         Decode the DF11 relies of the aviation obstacles in the $m$th beam space}
    }{
      Go to step 2\;
    }
    $N_{\rm av} \leftarrow$   Number of detected aviation obstacles\;
    \For{$k=1,2,\ldots,N_{\rm av}$}{
        Send the UF4 roll-call Mode S interrogation for the $k$th aviation obstacle and wait for its DF4 reply\;
        Measure the vector ${\bf y}$ and the matrix ${\bf Y}$  using the omnidirectional and the antenna array\;
        Use ${\bf y}$ in \eqref{to_estimate} to estimate $\hat{n}_k$\;
        Use ${\bf y}$ and $\hat{n}_k$ in \eqref{decode_ppm} to obtain $\hat{{\bf{b}}}_k$\;
        Employ $\hat{{\bf{b}}}_k$ in \eqref{set} and obtain $\hat{\mathcal I}=\{\hat{i}_0,\hat{i}_1,\ldots,\hat{i}_{i_{55}}\}$\;
       Downsample ${\bf y}$ as ${x}_i \triangleq  y_{\hat{n}_k+16M+[M / 2]+iM}$ for
       $i \in \hat{\mathcal I}$\;
       Obtain $\hat{\bm \varphi} \triangleq \big{[} \measuredangle x_{\hat{i}_0}  \ \measuredangle x_{\hat{i}_1} \ \ldots \ \measuredangle x_{\hat{i}_{55}}\big{]^{\rm T}}$\;
       Use $\hat{\bm \varphi}$ and $\hat{\bm u} \triangleq \big{[} {\hat{i}_0}  \ {\hat{i}_1} \ \ldots \ {\hat{i}_{55}}\big{]}^{\rm T}$ in \eqref{estimator} and obtain $\hat{v}_k$\;
      Flag$=$CRC$(\hat{{\bf{b}}}_k)$\;
       Use $\bf Y$, $\hat{n}_k$, and $\hat{\mathcal I}$ to obtain ${\bf Y}^{\rm a}$ in \eqref{array}\;
       Obtain $\hat{\bf A}=\frac{1}{56M}({\bf Y}^{\rm a})^{\rm H}{\bf Y}^{\rm a}$\;
       Use the eigenvectors of $\hat{\bf A}$ in \eqref{optimization2} and then solve the maximization problem in \eqref{optimization} to estimate $\hat{\theta}_k$ and $\hat{\phi}_k$\;
       \eIf{$\text{Flag}=1$}{
       $\hat{h}_k=\text{Decode}(\hat{{\bf{b}}}_k)$\;
       \eIf{$|\hat{\theta}_k| > \epsilon_{\theta}$}{
       $\hat{R}_k = ({\hat{h}_k-h})/{\sin(\hat{\theta}_k)}$\;
       }{
       $\hat{R}_k = \frac{cT_{\rm s}\hat{n}_k}{2}$\;
       }
       }{
       $\hat{R}_k = \frac{cT_{\rm s}\hat{n}_k}{2}$\;
       }
       Use $[\hat{R}_k \ \hat{\theta}_k \  \hat{\phi}_k \ \hat{v}_k]^{\rm T}$ as the input of intruder identification in Table II  and assess the risk status of the $k$th aviation obstacle $S_k$\;
       $\mathcal{R}\leftarrow\mathcal{R} \cup S_k$\;
       $\mathcal{P}\leftarrow\mathcal{P} \cup [\hat{R}_k \ \hat{\theta}_k \  \hat{\phi}_k \ \hat{v}_k]^{\rm T}$\;
       }
       Return $\mathcal{S}$ and  $\mathcal{P}$     
  \caption{SAA for $N_{\rm av}$ aviation obstacles}\label{algorithmrt}
\end{algorithm}

\section{Simulation}
In this section, we evaluate the performance of the proposed estimators and the SAA system under different scenarios.

\subsection{Simulation Setup}
Unless otherwise mentioned, the peak power of the DF4 reply by the Mode S transponder of each aviation obstacle is considered to be $8$ dBW \cite{Jeff2007}, and it is assumed that the UAV uses two planar antenna arrays with $(N_{\rm x}=2) \times (N_{\rm y}=2)$ elements, spaced by $d_{\rm x}=d_{\rm y}=\lambda/2  = 0.1375$ m in rows and columns,
to estimate the azimuth and elevation angles of the aviation obstacles. The elements of the array are backbaffled omnidirectional.
The  3dB beamwidth of the array for the azimuth and elevation angles is $\Delta \phi_3=\Delta \theta_3 \approx  60^{\circ}$, and 
the total number of scans for UF11 Mode S interrogation to cover $\phi \in (0, 180)$ and $\theta \in (-90, 90)$ is $M_{\rm e}M_{\rm a}=9$.
The raise and decay time of the Mode S trapezoidal transmit pulse in \eqref{trap} is $\tau_{\rm r}=0.01\mu $s, $T=0.5\mu$s \cite{leonardi2017air}, and $\tau_{\rm gr}=270 \mu s$. 
The bandwidth of the low-pass filter of root-raised cosine with roll-off factor $\beta=0.95$ and unit energy $E_{\rm h}=1$ at the omnidirectional antenna and the antenna arrays is $B=6$ MHz \cite{WinNT}.
The sampling time at the UAV receiver is $T_{\rm s}= (1+\beta) / (2B)$, which results in $6$ samples per PPM symbol and $M= [T/T_{\rm s}]=3$. The PSD of the noise $N_0$ over the filter bandwidth, i.e., $f \in [-B,B]$ is $N_0=2.4\times 10^{-21}$ {watts per hertz}.
The maximum time delay to receive the DF4 reply after the transmission of the UF4 Mode S interrogation is $\tau_{\max}= 0.0214$ s.
The performance of the proposed estimators is evaluated in terms of RMSE for $10^4$ Monte Carlo trials.
The simulation setup parameters are summarized in Table \ref{tab:orbital_data}.

\begin{figure}[!t]
\centering
\includegraphics[width=3.2in]{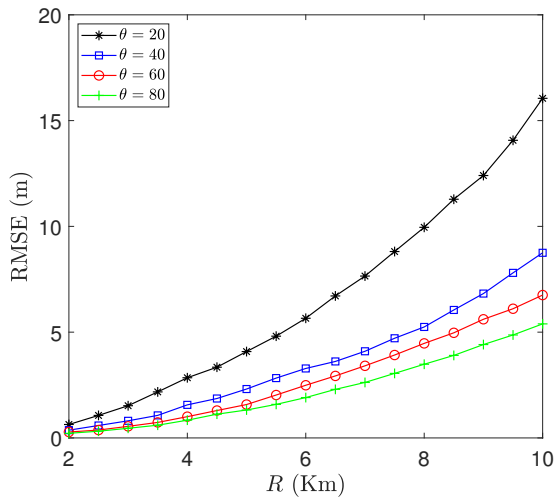}
\caption{The RMSE of the proposed ranging algorithm for different values of elevation angle $\theta$,  $\phi \in \mathcal{U}_{\rm c}(0,180)$, and $v_{\rm r} =55$ m/s.}\label{Mode_cccwq}
\end{figure}

\subsection{Simulation Results}
Fig.~\ref{Mode_cccwq} shows the RMSE of the proposed ranging algorithm versus the true range for different values of elevation angle $\theta$ for an aviation obstacle.
We consider that the azimuth angle and the relative radial velocity of the aviation obstacle and the UAV are distributed as $\phi \in \mathcal{U}_{\rm c}(0,180)$ and $v_{\rm r} =55$ m/s, respectively.
As expected, as the range increases, the RMSE of the proposed ranging algorithm increases. Also, the higher elevation angle $|\theta|$, the lower the RMSE of ranging because of the term $\sin(\hat{\theta})$ in the denominator of the estimator in \eqref{range}.
In terms of bias analysis, the simulation results show that the proposed ranging algorithm is unbiased for $ 0<R \leq 10$ km.

\begin{figure}[!t]
\centering
\includegraphics[width=3.2in]{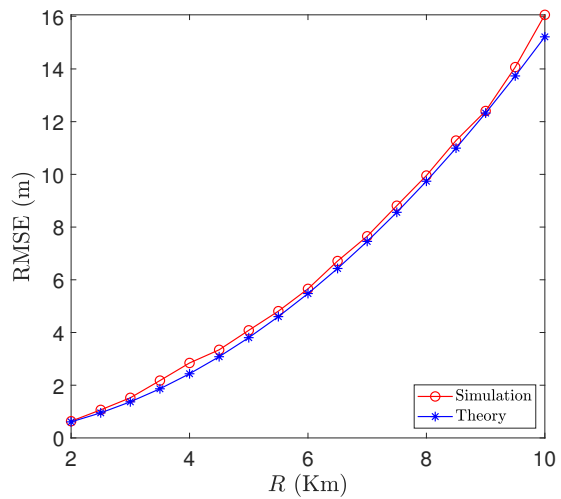}
\caption{Theoretical RMSE versus empirical RMSE of the proposed ranging algorithm for $\theta=20^{\circ}$.}\label{Mode_cccwqwe}
\end{figure}

In Fig.~\ref{Mode_cccwqwe}, we compare the empirical RMSE of the proposed ranging algorithm obtained by simulation experiments and the theoretical RMSE derived in \eqref{var_range} for different values of $R$ and $\theta=20^{\circ}$. As seen, there is a small gap between
the empirical RMSE and the theoretical one. 
The gap between the empirical and the analytical RMSE decreases as the number of PPM symbols increases by receiving multiple DF4 replies or by using extended-length DF4 downlink replies.

\begin{figure}[!t]
\centering
\includegraphics[width=3.2in]{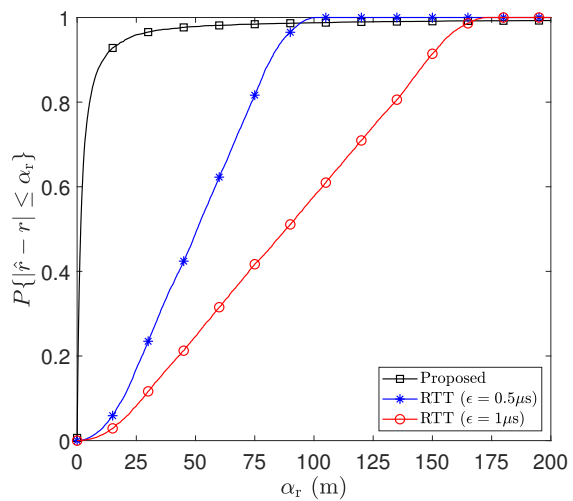}
\caption{Range error probability comparison of the proposed ranging algorithm with the RTT-based ranging algorithm.}\label{probability}
\end{figure}

Fig.~\ref{probability} compares the range error probability of the RTT-based and the proposed ranging algorithms.
The range error probability is defined as $\mathbb{P}\{|\hat{r}-r|\leq \alpha_{\rm r}\}$, and it is assumed that the azimuth and elevation angles of the aviation obstacle are $\phi \in \mathcal{U}_{\rm c}(0,180)$ and $|\theta| \in  \mathcal{U}_{\rm c}(1,90)$, respectively, and the relative radial velocity of the UAV is $v_{\rm r} =55$ m/s. For the RTT-based ranging algorithm, we consider that there is $r_\epsilon \sim \mathcal{U}_{\rm c}(0,\epsilon)$ random delay in the response of the Mode S transponder. 
As seen, the proposed ranging algorithm outperforms the RTT-based ranging algorithm even for small values of $\epsilon$.
\begin{figure}[!t]
\centering
\includegraphics[width=3.2in]{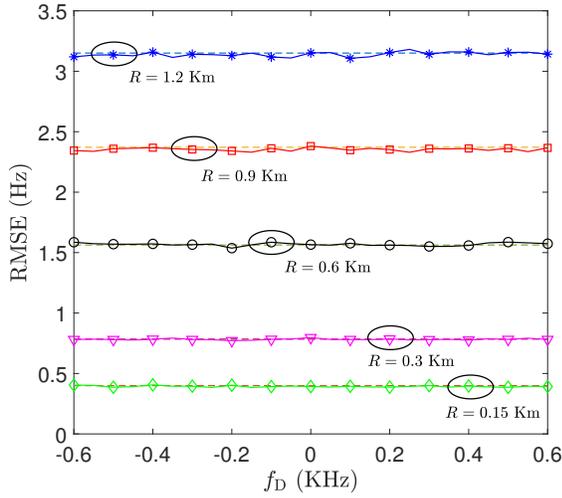}
\caption{The RMSE of the proposed Doppler frequency shift estimator for different values of $R$. The dashed line shows the theoretical RMSE.}\label{Mode_dop}
\end{figure}
\begin{figure}[!t]
\centering
\includegraphics[width=3.2in]{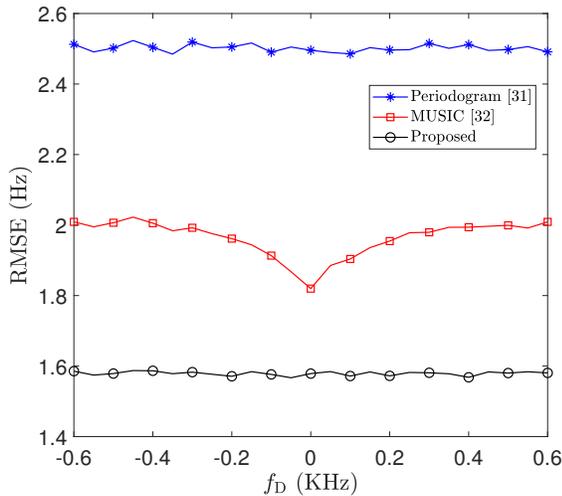}
\caption{Performance comparison of the proposed Doppler frequency shift estimator with the periodogram and the MUSIC estimators.}\label{Doppler_compare}
\end{figure}

Fig.~\ref{Mode_dop} illustrates the RMSE of the proposed Doppler frequency shift estimator for $-600 \leq  f_{\rm D}\leq 600$ Hz
and different values of the range $R$.
We consider that the elevation and azimuth angles of the aviation obstacle are distributed as $\phi \in \mathcal{U}_{\rm c}(0,180)$ and  $|\theta| \in  \mathcal{U}_{\rm c}(1,90)$.
We also show the theoretical RMSE derived in \eqref{var}. As seen, the proposed estimator offers low RMSE at high SNR values. Moreover, the RMSE obtained by simulation experiments matches the theoretical RMSE.

In Fig.~\ref{Doppler_compare}, we compare the RMSE of the proposed Doppler frequency shift estimator with the periodogram estimator in \cite{fulop2006algorithms} and the MUSIC estimator in  \cite{stoica2005spectral}. The azimuth and elevation angles of the aviation obstacle are assumed to be $\phi \in \mathcal{U}_{\rm c}(0,180)$ and $|\theta| \in  \mathcal{U}_{\rm c}(1,90)$, respectively, and the range between the UAV and the aviation obstacle is $R=0.6$ Km.
As seen, the proposed Doppler frequency shift estimator outperforms the MUSIC and the periodogram estimators
with lower computational complexity. {The complexity order of the proposed Doppler frequency shift estimator, the Lomb-Scargle periodogram, and the MUSIC are $\mathcal{O}(N_{\rm e}^2)$, $\mathcal{O}(N_{\rm e}F)$ \cite{lomb1976least}, and $\mathcal{O}(N_{\rm a}^3+N_{\rm a}^2N_{\rm e}+F)$ \cite{gentilho2020direction}, respectively, where $N_{\rm a}$ is the number of array elements, $F$ is the number of frequency grids, and $N_{\rm e}$ is the number of observation samples used for estimation.}

\begin{figure}[!t]
\vspace{2em}
\centering
\includegraphics[width=3.2in]{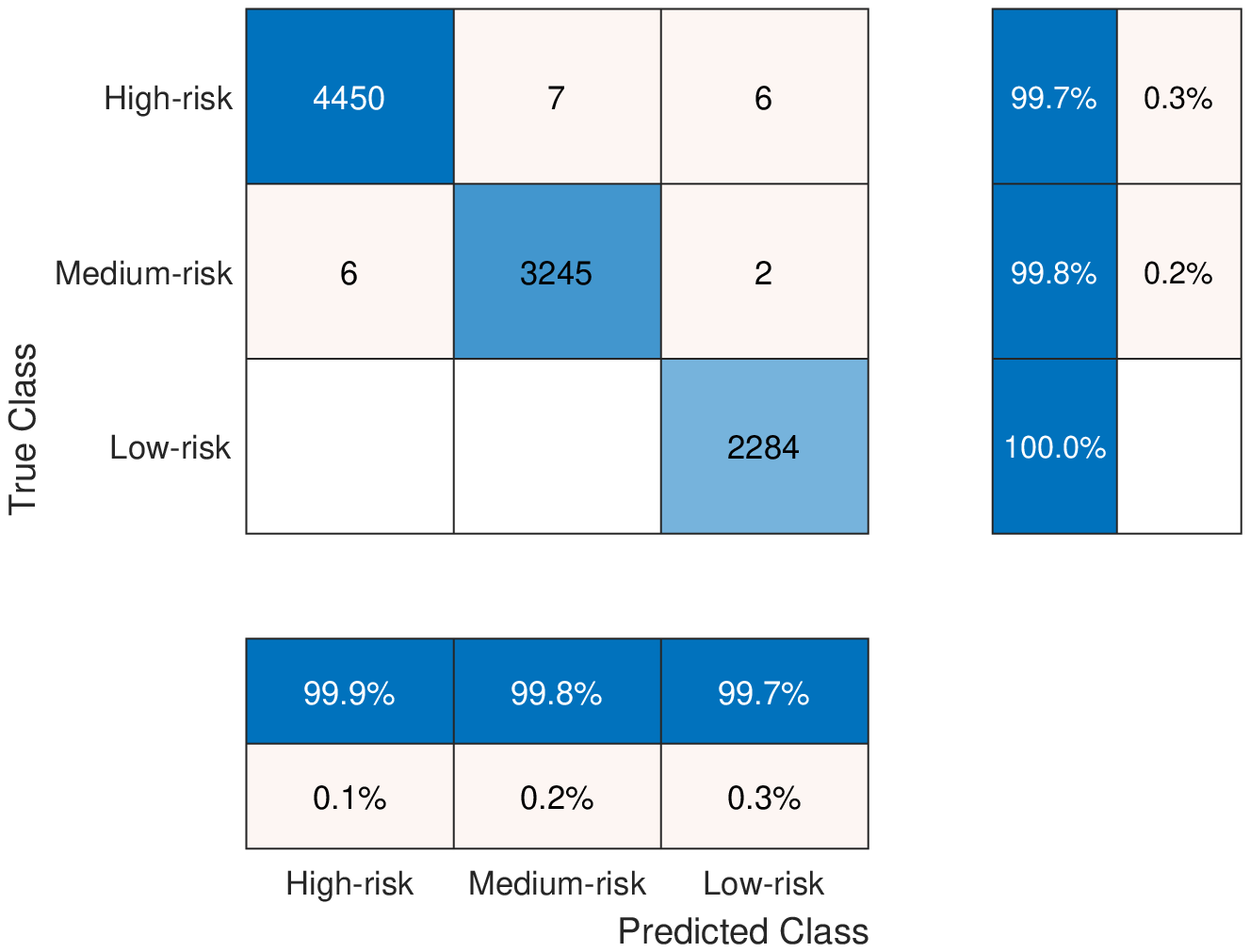}
\caption{Performance of the proposed intruder identification method for $10^4$ Monte Carlo trials.}\label{Mode_cccwqweE3WE}
\end{figure}

In Fig.~\ref{Mode_cccwqweE3WE}, we evaluate the performance of the intruder identification method developed using the proposed range and radial velocity estimators. Without loss of generality, we consider that the movement direction of the UAV is $\phi_{\rm v}= 0^{\circ}$ and $\theta_{\rm v}=0^{\circ}$ in the coordinates of the array.
For 8 aviation obstacles in the airspace, we assume that  $\phi \in \mathcal{U}_{\rm c}(0,180)$, $|\theta| \in  \mathcal{U}_{\rm c}(1,90)$, and
$R \in \mathcal{U}_{\rm c}(0.1,1)$ km. 
The relative radial velocity of the aviation obstacle and the UAV is assumed to be $v_{\rm r} \in \mathcal{U}_{\rm c}(-139,+139)$ m/s. The safety thresholds in Table II are considered to be
$\lambda_R=0.5$ km, $\lambda_{\phi}=10^{\circ}$, $\lambda_{\theta}=10^{\circ}$, and $\lambda_v=20$ m/s. 
The safety parameter associated with the radial velocity
$\lambda_v=20$ m/s  gives  $25$ s time to the UAV to avoid any possible collision with the aviation obstacle.
As seen  in Fig.~\ref{Mode_cccwqweE3WE}, the proposed intruder identification mechanism can accurately classify the aviation obstacles.
High-risk intruders can be identified with a probability of 0.997, which is significant.

\section{Conclusions}
A cooperative RF-based SAA algorithm for the safe navigation of small-sized UAVs was proposed in this paper. The proposed SAA algorithm takes the advantage of the Mode S SSR like the ACAS; however, it is capable of radial velocity estimation using a single DF4 reply. In addition, it uses a new ranging algorithm based on the triangular relation between the estimated elevation angle and the altitude difference between the aviation obstacle and the UAV. The proposed triangulation-based ranging algorithm is more accurate compared to the RTT-based ranging algorithm, and their combination is a promising solution for high accuracy ranging.
The proposed SAA system benefits from an intruder identification mechanism  that can accurately identify the risk level of the aviation obstacles in the airspace.
Future work concerns the applicability of the proposed SAA for industrial aircraft, the evaluation of the proposed SAA with real data, the development of channel access for the simultaneous detection of multiple UAVs, and the combination of 5G technologies and SSR.

\appendices
\section{}\label{Appx_A}
Let $n_0$ denote the index of the first sample of the DF4 reply received in the UAV omnidirectional antenna receiver. The index of the sample denoting the start of the payload symbols is $n_{\rm s} \triangleq n_0+16M$.
We can write the $K \triangleq 112M$ observation samples corresponding to the payload PPM vector $\bar{\bf b} = {[}\bar{b}_0 \ \bar{b}_1 \ \ldots \  \bar{b}_{111}  {]}^{\rm T} = {[}b_0 \ \neg{b}_0 \ b_1 \ \neg{b}_1 \ \ldots \  b_{55} \ \neg{b}_{55} {]}^{\rm T}$,
i.e., $y_{n_0+16M}, y_{n_0+16M+1},\ldots,y_{n_0+16M+K-1}$ as
\begin{align}\label{synch}
\hspace{-1.5em} {\bf a} =  \big{[}a_0 \  a_1 \ \ldots \  a_{K-1} \big{]}^{\rm T}
\approx \frac{\sqrt{P_{\rm t}\eta_0}}{L} {\bf F}_{\rm s}(\bar{\bf b}\otimes {\bf g})+{\bf v}_{\rm s},
\end{align}
where $a_k \triangleq y_{n_0+16M+k}$, ${\bf F}_{\rm s} \triangleq {\rm diag} ({\bf f}_{\rm s})$,
\begin{align}
{\bf f}_{\rm s} \triangleq & \big{[}e^{j(2\pi f_{\rm D}T_{\rm s}n_{\rm s}+\psi_0)} \ e^{j(2\pi f_{\rm D}T_{\rm s}(n_{\rm s}+1)+\psi_0)}
\\ \nonumber
& \,\,\,\,\,\,\,\,\,\,\,\,\,\,\,\,\,\,\,\,\,\,\,\,\,\,\,\,\,\,\,\,\,\,\,\,\,\,\,\ \dots \
e^{j(2\pi f_{\rm D}T_{\rm s}(K+n_{\rm s}-1)+\psi_0)}
\big{]}^{\rm T},
\end{align}
$
{\bf v}_{\rm s} \triangleq  {[}w_{n_{\rm s}} \ w_{n_{\rm s}+1} \ \ldots \ w_{n_{\rm s}+K-1} {]}^{\rm T}
$,
and 
${\bf{g}}=[g_0 \ g_1 \ \ldots \  g_{M-1}]^{\rm T}$.
To estimate the Doppler frequency shift, we can use the $M$-fold decimation of $\bf{a}$ as
\begin{align} \nonumber
{\bf x} & =\big{[}x_{0} \  x_{1} \ \ldots \  x_{111} \big{]}^{\rm T}
 =\big{[}a_{n_{\rm c}} \  a_{n_{\rm c}+M} \ \ldots \  a_{n_{\rm c}+111M} \big{]}^{\rm T} \\  \label{down_sample}
& \approx \frac{\sqrt{P_{\rm t}\eta_0}}{L} g_{n_{\rm c}} {\bf F}_{\rm d}\bar{\bf b}+{\bf v}_{\rm d},
\end{align}
where $n_{\rm c} \triangleq [M / 2]$, ${\bf F}_{\rm d} \triangleq {\rm diag} ({\bf f}_{\rm d})$,
\begin{align}
{\bf f}_{\rm d} \triangleq & \big{[}e^{j(2\pi f_{\rm D}T_{\rm s}n_{\rm d}+\psi_0)} \ e^{j(2\pi f_{\rm D}T_{\rm s}(n_{\rm d}+M)+\psi_0)}
\\ \nonumber
& \,\,\,\,\,\,\,\,\,\,\,\,\,\,\,\,\,\,\,\,\,\,\,\,\,\,\,\,\,\,\,\,\,\,\,\,\,\,\,\ \dots \
e^{j(2\pi f_{\rm D}T_{\rm s}(n_{\rm d}+111M)+\psi_0)}
\big{]}^{\rm T},
\end{align}
and
${\bf v}_{\rm d} \triangleq \big{[}w_{n_{\rm d}} \  w_{n_{\rm d}+M} \ \ldots \  w_{n_{\rm d}+111M} \big{]}^{\rm T}$,
with $n_{\rm d} \triangleq n_{\rm s}+n_{\rm c}=n_0+16M+[M / 2]$.
$g_{n_{\rm c}}$, $n_{\rm c} \triangleq [M / 2]$, is the $(n_{\rm c}+1)$th elements of the pulse shaping vector ${\bf g}$ given in \eqref{guiio21}.
The vector ${\bf x}$ is obtained by sampling in the middle of the received pulses.

At moderately high SNR values
that is the case in SAA systems, the signal model in \eqref{down_sample} can be approximated by \cite{tretter1985estimating}
\begin{align}\nonumber
{x}_i & = y_{n_{\rm d}+iM}  = y_{n_0+M(16+i)+n_{\rm c}}=  y_{n_0+M(16+i)+[\frac{M}{2}]}\\ \label{moderate_snr}
& \approx  \frac{\sqrt{P_{\rm t}\eta_0}}{L} g_{n_{\rm c}} e^{{j(2\pi f_{\rm D}T_{\rm s} (n_{\rm d}+iM)+\psi_{\rm 0}+\vartheta_i)}},\,\,\,\,\,\,\  i \in {\mathcal I},
\end{align}
where the set ${\mathcal I}$ with cardinality $|{\mathcal I}|=56$ is given in \eqref{set}, $\vartheta_i\in \mathbb{R}$, $i \in {\mathcal I}$,  is zero-mean white Gaussian noise with variance  $\sigma_{\rm \vartheta}^2=\frac{1}{2} (L^2 \sigma_{\rm w}^2) / (P_{\rm t}\eta_{\rm 0}g_{n_{\rm c}})$.
The phase of ${x}_i$ can be written as
\begin{align}\label{phase}
\measuredangle x_{i} &= \measuredangle y_{n_{\rm d}+iM} \\ \nonumber
& \approx  2\pi f_{\rm D}T_{\rm s} (n_{\rm d}+iM)+\psi_{\rm 0}+\vartheta_i,\,\,\,\,\,\,\  i \in {\mathcal I}.
\end{align}
The Doppler frequency shift between the UAV and the aviation obstacle $f_{\rm D}$ can be estimated by the phase of the received signal in \eqref{phase}.
Let us write the phase difference of the received signal in \eqref{phase} in vector form as
\begin{align}\label{linear_model}
{\bm \Delta}{\bm \varphi} =2\pi M f_{\rm D}T_{\rm s}{\bm \Delta}{\bf u}  +{\bm \Delta}{\bm \vartheta},
\end{align}
where
$
{\bm \varphi} \triangleq \big{[} \measuredangle x_{i_0}  \ \measuredangle x_{i_1} \ \ldots \ \measuredangle x_{i_{55}}\big{]}^{\rm T},
$
$
{\bm u} \triangleq \big{[} {i_0}  \ {i_1} \ \ldots \ {i_{55}}\big{]}^{\rm T},
$
$
{\bm \vartheta} \triangleq \big{[} \vartheta_{i_0}  \ \vartheta_{i_1} \ \ldots \ \vartheta_{i_{55}}\big{]}^{\rm T},
$
$i_k \in {\mathcal I}$ and $i_0 < i_1 < \ldots < i_{55}$, and
 ${\bm \Delta }$ is the $55\times 56$ discrete derivative matrix given in \eqref{derivative}.
We can easily show that for PPM signaling, we have $i_{k+1}-i_k \in \{1,2,3\}$ for $i_k\in {\mathcal I}$.

Because ${\bm \vartheta} \sim \mathcal{N}(0,\sigma_{\vartheta}^2{\bf I})$, the noise vector ${\bm \Delta}{\bm \vartheta} = {[} \vartheta_{i_0}-\vartheta_{i_1}  \ \vartheta_{i_1}-\vartheta_{i_2} \ \ldots \ \vartheta_{i_{54}}-\vartheta_{i_{55}}]^{\rm T}$ in \eqref{linear_model} is zero-mean and colored with covariance matrix $\bf{C}$, where
the entry on the $m$th row and $n$th column of the covariance $\bf{C}$, i.e.,
$C_{mn}=\mathbb{E}\{(\vartheta_{i_{(m-1)}}-\vartheta_{i_m})(\vartheta_{i_{(n-1)}}-\vartheta_{i_n})\}$ is given by
\begin{align}
{\bf{C}}_{mn} = \sigma_{\vartheta}^2 \left\{ \begin{array}{l}
2 \,\,\,\,\,\,\,\,\,\,\,\,\,\,\,\,\,\,\,\,\,\,\,\,\,\,\,\,\,\,\,\  m=n\\
-1 \,\,\,\,\,\,\,\,\,\,\,\,\,\,\,\,\,\,\,\,\,\,\,\,\,\,\,\  m \in \{n-1,n+1\} \\
0 \,\,\,\,\,\,\,\,\,\,\,\,\,\,\,\,\,\,\,\,\,\,\,\,\,\,\,\,\,\,\,\  {\rm otherwise}. \\
\end{array} \right.
\end{align}
Using (7-43) and (7-44) in \cite{kay1993fundamentals}, the optimal maximum likelihood estimation (MLE) of the Doppler frequency shift, $f_{\rm D}$, and its RMSE for the linear observation model in \eqref{linear_model} are obtained as in
\eqref{estimator} and \eqref{var}, respectively.
Since $\mathbb{E}\{{\bm \Delta}{\bm \vartheta}\}={\bf 0}_{55}$, we have $\mathbb{E}\{{\bm \Delta}{\bm \varphi}\} =2\pi M f_{\rm D}T_{\rm s}{\bm \Delta}{\bf u}$, and thus, we can write
\begin{align}\label{unbiased}
 \mathbb{E} \{\hat{f}_{\rm D}\}  =\frac{1}{2\pi M T_{\rm s}} \frac{{\bf u}^{\rm T} {\bm \Delta}^{\rm{T}} {\bf C}^{-1}\mathbb{E}\{{\bm \Delta} {\bm \varphi}\}}{{\bf u}^{\rm T}{\bm \Delta}^{\rm T} {\bf C}^{-1}{\bm \Delta}{\bf u}}={f}_{\rm D},
\end{align}
which shows that the proposed Doppler frequency shift estimator is an unbiased estimator.

\section{}\label{Appx_C}
Let the sequence $\{\text{PRI}_1,\text{PRI}_2,\ldots,\text{PRI}_{Q}\}$, $Q\in \mathbb{N}$, denote the PRI group of a pulse radar, where ${\text{PRI}}_i\triangleq \frac{a_i}{b_i}$, and ${a_i}$ and ${b_i}$ are coprime integers. 
The maximum unambiguous Doppler frequency shift of this waveform is given by \cite{lu2016maximum}
\begin{align}\label{amb_eq}
|f_{\rm D}| \leq \frac{\text{LCM}(b_1,b_2,\ldots,b_{Q})}{\text{GCD}(a_1,a_2,\ldots,a_{Q})},
\end{align}
where LCM and GCD denote the least common multiple and the greatest common devisor, respectively. 
For the DF4 PPM waveform with $T=0.5\mu {\rm s}$, we have ${\text{PRI}}_i\triangleq \frac{a_i}{b_i}\in \{T,2T,3T\}=\big{\{}\frac{1}{2\times 10^6},\frac{1}{10^6},\frac{3}{2\times 10^6}\}$.
 Hence, we can write 
\begin{align}\label{doppler_up}
|f_{\rm D}| & = \leq \frac{\text{LCM}(b_1,b_2,\ldots,b_{Q})}{\text{GCD}(a_1,a_2,\ldots,a_{Q})} \\ \nonumber 
& =\frac{\text{LCM}(2\times 10^6, 10^6,2\times 10^6)}{\text{GCD}(1,1,3)}=2\times 10^6=\frac{1}{T}.
\end{align} 
Finally, by substituting \eqref{doppler_up} into $|f_{\rm D}| = \frac{|v_{\rm r}|}{\lambda}$, we obtain \eqref{xxy}.

\section{}\label{Appx:d}
Here, we exploit the results in  \cite{stoica1989music}  to analytically obtain
the asymptotic MSE of the 2D-MUSIC for the planar antenna array used in the proposed SAA system.
By defining $\omega_{\theta} \triangleq \sin(\theta)$, $\cos(\theta)=\sqrt{1-\omega_{\theta}^2}$, and exploiting the result in (3.12) of \cite{stoica1989music}, we asymptotically obtain
the 2D-MUSIC estimation error of ${\hat{\omega}}_{\theta} =  \sin(\hat{\theta})$ as
\begin{align}\label{el_var12}
\mathbb{E}\big{\{}({\hat{\omega}}_{\theta}-{{\omega}}_{\theta})^2\big{\}} = \frac{\frac{\sigma_{\rm w}^2}{112M}\frac{\lambda_{\rm max}}{(\sigma_{\rm w}^2-\lambda_{\rm max})^2}\big{|}{\bf a}^{\rm H}(\theta,\phi){\bf{e}}_1\big{|}^2}{\sum_{n=2}^{N_{\rm a}}\big{|}{\bm \alpha}^{\rm H}({\theta},\phi){\bf e}_n\big{|}^2},
\end{align}
where ${\bm \alpha}({\theta},\phi)\triangleq \big{[}\alpha_{1,1}({\theta},\phi) \ \alpha_{2,1}({\theta},\phi) \ \ldots \ \alpha_{N_{\rm x},N_{\rm y}}({\theta},\phi)\big{]}^{\rm T}$ with $\alpha_{m,n}({\theta},\phi)$ in \eqref{qqqty}, and $a_{m,n}({\theta},\phi)$ is given in \eqref{manifold_op}. In \eqref{el_var12},
the vector ${{\bf{e}}}_1$ is the eigenvector corresponding to the largest eigenvalue of $\Lambda \triangleq \xi{\bf a}^{*}(\theta,\phi){\bf a}^{\rm T}(\theta,\phi)+\sigma_{\rm w}^2{\bf I}$, i.e., $\lambda_{\rm max}$,  ${{\bf{e}}}_k$, $k=2,3,\ldots,N_{\rm x}N_{\rm y}$ are the eigenvectors corresponding to the $N_{\rm x}N_{\rm y}-1$ eigenvalues $\lambda=\sigma_{\rm w}^2$, and $\xi$ is given in \eqref{eq:p}.

Similarly, by defining $\omega_{\phi} \triangleq \sin(\phi)$, and exploiting the result in (3.12) of \cite{stoica1989music}, we obtain
the asymptotic 2D-MUSIC estimation error for ${\hat{\omega}}_{\phi} = \sin(\hat{\phi})$ as
\begin{align}\label{el_var}
\mathbb{E}\big{\{}({\hat{\omega}}_{\phi}-{{\omega}}_{\phi})^2\big{\}} = \frac{\frac{\sigma_{\rm w}^2}{112M}\frac{\lambda_{\rm max}}{(\sigma_{\rm w}-\lambda_{\rm max})^2}\big{|}{\bf a}^{\rm H}(\theta,\phi){\bf{e}}_1\big{|}^2}{\sum_{n=2}^{N_{\rm a}}\big{|}{\bm \beta}^{\rm H}({\theta},\phi){\bf e}_n\big{|}^2},
\end{align}
where ${\bm \beta}({\theta},\phi)\triangleq \big{[}\beta_{1,1}({\theta},\phi) \ \beta_{2,1}({\theta},\phi) \ \ldots \ \beta_{N_{\rm x},N_{\rm y}}({\theta},\phi)\big{]}^{\rm T}$
with $\beta_{m,n}({\theta},\phi)$ as
\begin{align}
\beta_{m,n}({\theta},\phi) &\triangleq \frac{\partial a_{m,n}({\theta},\phi)}{\partial \sin(\phi)} \\ \nonumber
& = \frac{j2\pi}{\lambda} d_{\rm y}(n-1)\sin(\theta)a_{m,n}({\theta},\phi).
\end{align}
Similar to \cite{stoica1989music}, we can show that $\hat{\omega}_{\theta}-{\omega}_{\theta}$ and $\hat{\omega}_{\phi}-{\omega}_{\phi}$  are asymptotically jointly Gaussian distribution with zero-mean.

\section{}\label{Appx_B}
The authors in \cite{stoica1989music} proved that  $\hat{\omega}_{\theta}$ is asymptotically an unbiased estimate of ${\omega}_{\theta}={\sin}({\theta})$.  Thus, we can write $\hat{\omega}_{\theta} =   {\sin}(\hat{\theta}) =\sin(\theta)+Z$,
where $Z \sim \mathcal{N}(0,\mu)$, and $\mu \triangleq  \mathbb{E}\big{\{}({\hat{\omega}}_{\theta}-{{\omega}}_{\theta})^2\big{\}}$ is given in \eqref{el_var12}.
By defining $X\triangleq Z / {\omega}_{\theta}$ (for moderately high SNR values and sufficiently large $|\epsilon_{\theta}|$, we have $\mathbb{P}\{|X|<1\} \approx 1$) and using the series representations $\frac{1}{1+X} = \sum_{n\geq 0} (-1)^n X^n$, $|X|<1$,
for $|\theta| >\epsilon_{\theta}$, we can write
\begin{align}\label{taylor}
\hat{R} &= \frac{h_{\rm a}-h}{\sin(\hat{\theta})} = \frac{h_{\rm a}-h}{\sin(\theta)+Z}
\\ \nonumber
& =(h_{\rm a}-h)\sum_{n \geq 0}(-1)^n \frac{Z^n}{(\sin(\theta))^{n+1}}
\\ \nonumber
& =\frac{(h_{\rm a}-h)}{\sin(\theta)} \Bigg{(}1-\frac{Z}{\sin(\theta)}+\mathcal{O}\Big{(}\frac{Z^2}{(\sin(\theta))^2}\Big{)} \Bigg{)}, \ \frac{Z}{\sin(\theta)}\rightarrow 0,
\end{align}
where $h$ and  $h_{\rm a}$ denote the altitude of the UAV and the aviation obstacle, respectively.
The series in \eqref{taylor} converges by the ratio test if $|Z|<|\sin(\theta)|$ and $|\theta| \geq \epsilon_{\theta}$.

For moderately high SNR values and $|\theta| \geq \epsilon_{\theta}$, $Z/ \sin(\theta)\rightarrow 0$; hence, by taking expectations on both sides of \eqref{taylor}, we obtain
\begin{align}
{\mathbb{E} \{\hat{R}\}} = \mathbb{E}\Bigg{\{}\frac{h_{\rm a}-h}{\sin(\hat{\theta})}\Bigg{\}}  \approx \frac{h_{\rm a}-h}{\sin({\theta})}.
\end{align}
The variance of $\{\hat{R}\}$ is also given by
\begin{align}
{{\rm var} \{\hat{R}}\} = {\rm var} \Bigg{\{}\frac{h_{\rm a}-h}{\sin(\hat{\theta})}\Bigg{\}}
\approx \frac{(h_{\rm a}-h)^2  {\rm var}\{Z\}  }{\sin^4({\theta})}.
\end{align}
Because ${\mathbb{E} \{\hat{R}\}} \approx R$, the MSE of the estimated range of the aviation obstacle is given by
\begin{align}\label{mse3}
{\mathbb{E} \{(\hat{R}-R)^2\}} \approx \frac{(h_{\rm a}-h)^2 {\rm var}\{Z\} }{\sin^4({\theta})}.
\end{align}
Finally, by replacing ${\rm var}\{Z\}$ with $\mathbb{E}\big{\{}({\hat{\omega}}_{\theta}-{{\omega}}_{\theta})^2\big{\}}$ in \eqref{el_var12},
 we obtain \eqref{var_range}.



\end{document}